\documentclass[sigconf]{acmart}

\AtBeginDocument{%
  \providecommand\BibTeX{{%
    \normalfont B\kern-0.5em{\scshape i\kern-0.25em b}\kern-0.8em\TeX}}}

\copyrightyear{2024}
\acmYear{2024}
\setcopyright{acmlicensed}\acmConference[CHI '24]{Proceedings of the CHI
Conference on Human Factors in Computing Systems}{May 11--16,
2024}{Honolulu, HI, USA}
\acmBooktitle{Proceedings of the CHI Conference on Human Factors in
Computing Systems (CHI '24), May 11--16, 2024, Honolulu, HI, USA}
\acmDOI{10.1145/3613904.3642771}
\acmISBN{979-8-4007-0330-0/24/05}

\renewenvironment{quote}
  {\list{}{\rightmargin=0.5cm \leftmargin=0.5cm}%
   \item\relax}
  {\endlist}

\usepackage{enumitem}

\usepackage{xcolor}
\usepackage{multirow}
\usepackage{array}
\usepackage{siunitx}
\usepackage{graphicx}

\usepackage{dblfloatfix}

\renewcommand{\paragraph}[1]{\vspace{0.2em}\noindent\textbf{\textit{#1}}\hspace*{.3em}}

\usepackage{xspace}

\usepackage{balance} 
\usepackage{hyperref} 
\usepackage{tabularx} 

\newcommand{\edits}[1]{#1}
\newcommand{\deletes}[1]{}
\newcommand{\movedto}[1]{#1}
\newcommand{\movedfrom}[1]{}

\begin{document}

\title{The Future of HCI-Policy Collaboration}
\settopmatter{authorsperrow=4}
\author{Qian Yang}
\affiliation{%
  \institution{Cornell University}
  \country{USA}
  }
\email{qianyang@cornell.edu}

\author{Richmond Y. Wong}
\affiliation{%
  \institution{Georgia Institute of Technology}
  \country{USA}
  }
\email{rwong34@gatech.edu}

\author{Steven J. Jackson}
\affiliation{%
  \institution{Cornell University}
  \country{USA}}
\email{sjj54@cornell.edu}

\author{Sabine Junginger}
\affiliation{%
  \institution{The Lucerne University of Applied Sciences and Arts}
  \country{Switzerland}}
\email{sabine.junginger@hslu.ch}

\author{Margaret Hagan}
\affiliation{%
  \institution{Stanford University}
  \country{USA}}
\email{mdhagan@stanford.edu}

\author{Thomas Krendl Gilbert}
\affiliation{%
  \institution{New York Academy of Sciences}
  \country{USA}
  }
\email{tgilbert@nyas.org}

\author{John Zimmerman}
\affiliation{%
  \institution{Carnegie Mellon University}
  \country{USA}}
\email{johnz@cs.cmu.edu}

\renewcommand{\shortauthors}{Yang et al.}

\begin{abstract}

Policies significantly shape computation's societal impact, a crucial HCI concern. However, challenges persist when HCI professionals attempt to integrate policy into their work or affect policy outcomes.
Prior research considered these challenges at the ``border'' of HCI and policy.
This paper asks: What if HCI considers policy integral to its intellectual concerns, placing system-people-policy interaction not at the border but nearer the center of HCI research, practice, and education?
What if HCI fosters a mosaic of methods and knowledge contributions that blend system, human, and policy expertise in various ways, just like HCI has done with blending system and human expertise? We present this re-imagined HCI-policy relationship as a provocation and highlight its usefulness: It spotlights previously overlooked system-people-policy interaction work in HCI. It unveils new opportunities for HCI's futuring, empirical, and design projects. It allows HCI to coordinate its diverse policy engagements, enhancing its collective impact on policy outcomes.

\end{abstract}




\maketitle


\section{Introduction}


From the future of work, to equity and sustainability, computation's societal impact has become a critical HCI concern. 
\movedto{At least three forces help shape this impact: the incentives, deterrents, and rules wired into computational systems (“system design”); the social and cultural tendencies of people (“social practice”); and the laws and regulations that govern both systems and people (“policy”)~\cite{policyknot-CSCW14,design-policy-workshop,Junginger2013-matters-of-design-in-policy,policy-generativity-Centivany-CSCW16}.}
Take privacy as an example: 
The design of a mobile app, its users' privacy awareness, and related policies (e.g., app stores' mandates, governments' regulatory fines) all help decide how well this app will protect user privacy~\cite{HongFATEPrivacyPolicy_CACM2023}. 

In many cases, HCI researchers and practitioners are already successful in improving system-people-policy interaction. For instance, HCI scholar Lorrie Cranor both conducted empirical and design research on privacy, and led the drafting of privacy regulations at the U.S. Federal Trade Commission (FTC)~\cite{OccupyCHI_USpolicy_CHI12panel}. Professor Colin Gray's work on deceptive web design patterns influenced the U.S. DETOUR (Deceptive Experiences To Online Users Reduction) Act~\cite{ftc_workshop_darkpattern,gray2022arizona_court}. 
\movedto{Such success stories span many countries and social issues, such as content moderation~\cite{haimson2021disproportionate}, gig worker protection~\cite{lee2015working,toxtli2021quantifying,newman2019found}, AI safety~\cite{LegalProvocationAutomation_NordiCHI22}, smart cities~\cite{HCITacticsForPolicyCHI21Whitney,disalvo2014making}, sustainability~\cite{blevis2007sustainable}, and more.} 
 
Yet in many other cases, policy considerations lag behind, or remain isolated from, the other HCI work that centers around systems and people. 
Consider Uber's creation of gig work (and intentional breach of taxi regulations) when it first launched its ride-sharing service~\cite{Uber_breaks_Law_INC15,Uber_breaks_Law_FastCompany13}, Facebook's adoption of AI newsfeed rankers (and amplification of misinformation)~\cite{haimson2021disproportionate}, or OpenAI's public release of chatGPT (and intentional violation of copyright laws)~\cite{TechCrunch2023_genAI_copoyright_lawsuits,Policybrief2023_GenAI_Jobs}.
Too often, HCI seemed to become excited about the technology first, take an interest in policy only after things had gone off the rails, and then find itself in the unenviable position of trying to put genies back into bottles. 

This paper aims to better understand this disparity and amplify HCI's collective voice in the policy realm. 
\movedto{From difficulties in bringing policymakers to the table~\cite{HCI-policy-boundary-CHI19},  to misalignment between HCI methods and policymakers' evidentiary needs~\cite{Spaa2022-evidence-and-policy}, many challenges of HCI-policy collaboration are well-known.}
\edits{Prior work referred to these challenges as at the ``\textit{boundaries}'' of HCI and policy~\cite{HCI-policy-boundary-CHI19}, prompting \movedto{decade-long calls for more collaborations across HCI-policy disciplinary boundaries~\cite{Lazar-interacting-with-public-policy,Lazar-HCI-policy-Interacitons15-forum,policyknot-CSCW14,OccupyCHI_USpolicy_CHI12panel,policy-generativity-Centivany-CSCW16,urquhart2017new,Junginger2013-matters-of-design-in-policy}.} 
These calls are necessary and valuable~\cite{Lazer_10year_HCIpulicpolicy2012}; nothing in this paper argues against them.
Meanwhile, the decade-long persistence of these calls also raises the question:}
\movedto{Can fresh perspectives on these challenges inspire additional strategies for overcoming them?}
\edits{This paper aims to provide one such perspective.}

\edits{This paper argues that---thanks to the decade-long calls for more HCI-policy collaboration---HCI has begun creating new methods and new types of knowledge contributions that are distinct from traditional HCI or policy ones. HCI and policy expertise have begun cross-pollinating. 
However, such unconventional, trans-disciplinary work has seemed to struggle with getting past reviewers, therefore has remained largely unrecognized in peer-reviewed HCI research venues, industry practices, or HCI textbooks.
This lack of recognition can contribute to the disparity in HCI's policy engagement and impact. }

\edits{
Reframing the challenges of HCI-policy collaboration as ``\textit{insufficient recognition of trans-disciplinary methods and knowledge}'' informs new strategies for addressing them:
\movedto{What if HCI were to see policy as integral to its intellectual concerns, placing system-people-policy interactions not ``\textit{at the boundaries of HCI and policy}''~\cite{HCI-policy-boundary-CHI19}, but nearer the \textit{center} of HCI research, practice, and education?}
What if HCI were to foster a mosaic of distinctively HCI methods and knowledge contributions that blended system, human, and policy expertise in varying ways and degrees, just like it has previously done with blending system and human expertise?

This paper sketches out this re-imagined HCI-policy landscape \textit{as a provocation}.
We demonstrate how it can (1) spotlight previously overlooked system-people-policy interaction work and knowledge, (2) unveil new opportunities for HCI's futuring, empirical, and design projects, and (3) allow HCI to coordinate its diverse policy engagements, thereby enhancing HCI's collective impact on policy outcomes.
We show how it opens up many new research questions, therefore is a useful seed for broader, HCI-community-wide discussions.
We invite fellow HCI researchers and practitioners to discuss their views of the relationship among systems, people, and policy expertise, and how HCI wants to position itself in this relationship.
}

This paper makes \edits{three}\deletes{two} contributions. 
First, it moves the research discourse on HCI-policy collaboration beyond the decade-long \textit{calls for more collaboration}, towards a community-wide discussion on \edits{its problem-solution framing.}\deletes{what ideal HCI-policy collaborations should look like.}
Second, it jump-starts this discussion by proposing one new problem-solution frame: There is insufficient recognition that policy is not just HCI's ``\textit{broader impact}'', but integral to HCI's intellectual pursuits in system-people interaction. Finally, it exposes important new questions for future research to address. For example, \edits{when is policy \textit{not} the best solution to tech's societal harms?}
\edits{And when sudden momentum for policy change appears, how might HCI quickly assemble its diverse work and seize the opportunity?}

\section{Key Concepts}\label{chapter:concepts}

At the outset of this paper, we plunged head first into the topics of ``\textit{policy}'' and ``\textit{HCI-policy collaboration}'' without detailing what they meant.
These are nuanced concepts, variably interpreted and lacking agreed-upon definitions~\cite{nudzor2009policy,junginger-book-USPS,van2016policy}. 
With these complexities in mind, we unpack what we mean by these terms. Our goal is not to establish conclusive definitions, but to scope them for the narrow purpose of this paper. 

\subsection{Policy}\label{chapter:def_policy}

In this paper, policy refers to principles, guidelines, and written rules formulated and adopted by an authority, organization, or government~\cite{policyknot-CSCW14}.
Among the widely-accepted policy definitions, we chose this one because it has both the specificity and breadth that match our research goals.

First, it encompasses various rules external to a computing system that wield authority over human-computer interactions, such as governmental policies, tech platform policies, and more.
All these policies help shape computation's societal impact, therefore hold relevance to HCI.

Secondly, this definition recognizes that a policy encompasses not only its written rules, but also the ways in which these rules are formulated and applied.
Take the EU AI Act as an example. Whose inputs shaped this legal framework~\cite{OpenAI_Lobbying_EU_AI}, how government agencies translate it into enforceable rules, and how the courts interpret and apply the rules to specific situations ~\cite{Brookings_EUAIAct_Enforcement2023} all influence AI's societal impact. All are relevant to HCI.

This definition of policy (and the scope of this paper) is also deliberately narrower than some prior work. \textit{It does not intend to address all aspects of policy practices, nor all political forces that can influence computation's societal impact.}
Because our goal is to understand the challenges of HCI-policy collaboration, we focus on the aspects of policy uncommon in current HCI work.
For example, a society's cultural politics (e.g., its conception of fairness~\cite{Fairness_Checklist}) and governmental policy-making procedures both significantly influence its tech policies. Yet because the former is already common in HCI work, the latter takes precedence in this paper.

\subsection{``\textit{Understanding}'' Policy, ``\textit{Designing}'' Policy}\label{chapter:def_understandDesignPolicy}
What this paper refers to as \textit{understanding} and \textit{designing} policy reflects this policy definition.
Because "policy" encompasses both the written rules and the process of their formulation and enactment, ``\textit{understanding}" a policy includes understanding both aspects.
Similarly, ``\textit{designing''} policy includes not only formulating rules, but also designing the processes for formulating and implementing the rules~\cite{ramesh2003studying}. In practice, policy design is the iterative process of (1) identifying policy needs, (2) clarifying policy needs (or issue-framing), (3) formulating policy, (4) designing systems and services that implement policy, and (5) evaluating policy outcomes~\cite{junginger-book-USPS,peters2021advanced}\footnote{This scope of policy ``\textit{design}'' is intentionally broader than some prior HCI literature, which focused on evidence-based policy-making (and HCI as a supplier of evidence)~\cite{HCI-policy-boundary-CHI19}.
It is also broader than some legal scholarship, which sees computing system ``\textit{design}'' as merely the implementation of a given policy requirement (e.g., generating a standard cookie opt-out menus~\cite{Tkacik_2020}, rather than a process of open-ended knowledge inquiry.}.
We chose a broader view of designing policy because we want to consider various HCI-policy connections~\cite{wong2019bringing,dorst2001creativity}.

The synergies between technology and policy design are self-evident. Both processes start with recognizing an opportunity for new systems/policies to help, and proceed with prototyping and evaluating their helpfulness iteratively.
These synergies have prompted some HCI researchers to call for designing technology and policy simultaneously~\cite{design-policy-workshop} and some legal scholars to call HCI a new direction in designing Information Technology law~\cite{urquhart2017new}.


\subsection{``\textit{Productive}'' HCI-Policy Collaboration}

By definition, HCI addresses the interactions between computing systems and people.
As these systems grew ubiquitous and integral to everyday life, HCI has expanded its focus from single-user-single-system interactions, to interactions among platforms, stakeholders, organizations, societal processes, and even planetary concerns~\cite{UCDdead,yangUXMLframework,blevis2007sustainable}.
\movedto{Policy entered HCI discourse during this expansion, because it significantly shapes these larger-scale and often political interactions~\cite{Lazer_10year_HCIpulicpolicy2012,policyknot-CSCW14,Dombrowski_SocialJusticeIXD_DIS16}.}

A productive HCI-policy collaboration can improve computation's societal impact, by catalyzing synergistic designs of computational systems, social practices, and policies~\cite{policyknot-CSCW14,pillowfort-platform-design-and-policy-chi22}.
Consider, for example, a novel AI system that helps diagnose a disease (system), clinician teams' diagnostic workflow (social practice), and clinical malpractice laws (policy).
To improve diagnostic accuracy in practice, designers cannot simplistically match clinician workflows and malpractice laws with this new AI system, nor vice versa. Instead, the three designs need to co-evolve~\cite{peters2021advanced}. Computation's societal impact improves step-wise, as the “knot” of system design, social practice design, and policy design loosens and tightens, unwinds, and reties~\cite{policyknot-CSCW14}.

There is no single best approach to HCI-policy collaboration that prescribes this ideal outcome~\cite{Jackson2007understanding_infra}.
Should HCI professionals generate empirical evidence of technological harm, before persuading policymakers to act~\cite{gray_darkpattern_CHI18}? Or should they envision a computational system and its regulations simultaneously~\cite{design-policy-workshop}? Should they seek a government position or work with advocacy groups when prompting new policies?
Depending on context, the most fruitful approach to HCI-policy collaboration varies, each bringing different challenges.

\subsection{Problem-Solution Frames}

How might we grasp the challenges of HCI-policy collaboration and devise solutions? This is a wicked problem~\cite{wicked_problems_rittel1973}. Because challenges facing every collaboration vary, there is no definitive answer to the question of ``\textit{which challenges(s) are responsible for the disparity in their success levels overall}'' or ``\textit{what solutions might help}''~\cite{Dombrowski_SocialJusticeIXD_DIS16}. 

Addressing wicked problems relies on problem framing and reframing~\cite{wicked_problems_rittel1973}. By seeing a problematic situation through different lenses, one can better understand its various dimensions and see new avenues for potential solutions~\cite{dorst2001creativity}. And \textit{this} is the goal of this paper. We wanted to find a new frame for understanding the challenges and disparity in HCI-policy collaboration; a frame that can reveal new avenues for solutions.


\section{Background}\label{chapter:bg}

HCI-policy collaboration has many well-known success stories and persistent challenges.
For example: difficulties in bringing policymakers to the table~\cite{HCI-policy-boundary-CHI19}, policymakers demanding policy evidence that (1) may not exist (e.g., evidence of a new technology's not-yet-manifested societal form) or (2) may not be the types of evidence HCI approaches produce~\cite{HCI-policy-boundary-CHI19,Spaa2022-evidence-and-policy}; mismatch between the pace of policy or political changes and the pace of HCI work ~\cite{OccupyCHI_USpolicy_CHI12panel,Nathan_Interactions2010_RwandaPolicy}; and more.

\edits{Interestingly, prior research rarely deliberated on the nature of these problematic situations. Instead, it directly suggested solutions.}
Most commonly, researchers called for more HCI-policy collaboration~\cite{Lazar-interacting-with-public-policy,Lazar-HCI-policy-Interacitons15-forum,policyknot-CSCW14,OccupyCHI_USpolicy_CHI12panel,policy-generativity-Centivany-CSCW16,urquhart2017new,Junginger2013-matters-of-design-in-policy}.
Others suggested more concrete solutions. For instance, to bring policymakers to the table, they encouraged more HCI scholars to take a gap year and work at government agencies~\cite{lazar2017_gap_year}. 
To address the HCI-policy evidentiary gap, they recommended involving policymakers in participatory design workshops, to help them appreciate HCI methods~\cite{HCI-policy-boundary-CHI19}.
To align the misaligned timelines of policy and HCI work, HCI 
researchers recommended committing to policy work for the long run and adjusting to the inherently slower pace of policy change~~\cite{OccupyCHI_USpolicy_CHI12panel}.
These recommendations yielded valuable results~\cite{Lazer_10year_HCIpulicpolicy2012}; nothing in this paper argues against them.
\edits{Nevertheless, we see opportunities to deliberate on the problem-solution frames that these solutions imply. Let us illustrate these opportunities through two examples.}

\subsection{Current Problem-Solution Frame \#1}\label{chapter:currentframe1_moreCollab}
The numerous calls for more HCI-policy collaboration suggest that a critical hindrance to this collaboration is that too few HCI professionals are participating~\cite{Lazar-interacting-with-public-policy}. 

However, the situation might be shifting.
\movedto{Over the past decade, methods such as participatory design and value-sensitive design---methods many policy actors also use---have moved more towards the center of HCI \cite{forlizzi2018moving,roto2021overlaps}. Since 2018, the ACM FAccT conference has been bringing together HCI, law and policy, and other fields to address AI's ethical issues~\cite{laufer2022four}.
Since 2021, SIGCHI publications mentioning “policy” surged by more than 40\%, according to ACM Digital Library} 
In this context, it is worth asking whether HCI's lack of attempts to policy engagement remains true today.


\subsection{Current Problem-Solution Frame \#2}\label{chapter:currentframe2_boundary}
HCI researchers have characterized the challenges of HCI-policy collaboration as occurring at the boundaries of HCI and policy~\cite{HCI-policy-boundary-CHI19}. This framing sees policy change as a ``\textit{broader impact}'' of HCI's intellectual work (i.e., understanding and designing system-people interaction). Therefore, the disciplinary boundaries between HCI and policy can hinder their collaboration.

Along this line, prior research repeatedly recommended that HCI and policy communities accept each other's norms.
For example, methodologically, HCI researchers recommended that policymakers embrace HCI's design methods~\cite{HCI-policy-boundary-CHI19}; socially, they encouraged more HCI professionals to socialize with policy actors~\cite{lazar2017_gap_year,Lazar-interacting-with-public-policy}; temporally, they encouraged HCI researchers to adapt to policy and political timelines~\cite{OccupyCHI_USpolicy_CHI12panel}.

These suggestions are highly valuable, but seem to be partial solutions for fostering productive HCI-policy partnerships. Take, for instance, the timeline of addressing the societal impact of generative AI (genAI). How should HCI and policy communities coordinate their pace of work to best address genAI's societal impact?
How should HCI communities approach publishing novel GPT applications, considering some \textit{might} be soon regulated out of existence?
These questions require nuanced debates~\cite{chatGPT_quickRegulation}. Urging one community to adopt the other's timeline appears insufficient.

More fundamentally, a productive collaboration between any two disciplines involves more than choosing whose norms to conform to. It entails creating new \textit{trans-disciplinary} methods and new bodies of knowledge distinct from either parent discipline~\cite{jantsch1972inter_transdisciplinary,klein1990interdisciplinarity}.
The innovations of computing systems, social practices, and policies require distinct methods, and each encompasses multiple temporal patterns~\cite{steinhardt2014reconciling,steinhardt2015anticipation,messeri2015greatest}.
Are there trans-disciplinary methods that can respect and bridge these differences? Such meta-discussions are absent in current HCI-policy research.

\edits{
\section{Research Activities}
\label{chapter:Method}
}

\movedto{We (a group of HCI, design, law, and policy researchers) 
wanted to reframe the challenges facing HCI-policy collaboration, in the hope of identifying new opportunities for addressing them.}
\edits{Towards this goal, we collected empirical reports of previous HCI-policy collaboration processes, analyzed factors contributing to their varying levels of success, and then worked to identify a new, useful problem-solution frame.}



\edits{
\subsection{Literature Review}
\label{chapter:method_litreview}

We first searched peer-reviewed HCI research publications and practitioner-facing books for empirical reports of previous and ongoing collaboration processes.}
\movedto{However, we soon realized that this literature has published very few such reports. 
Even the most fruitful collaborations---collaborations that resulted in national policies with HCI's scholars' names~\cite{OccupyCHI_USpolicy_CHI12panel,gray2022arizona_court}--- left little documentation of how HCI researchers and practitioners worked with policy actors in practice, or how they approached the gap between HCI and policy.}

HCI's Participatory Design (PD) projects are telling examples. Peer-reviewed HCI venues have published many such projects, where researchers investigated vulnerable stakeholders' needs and engaged them in drafting policy recommendations~\cite{young2019toward,HCITacticsForPolicyCHI21Whitney}.
While these publications offered detailed need-finding and rule-making processes, it is difficult to gauge whether or how they influenced policy outcomes. After all, citing academic references or attributing individual researchers is uncommon in many legal or public policy documents. 

\movedto{A larger set of HCI-policy collaboration efforts reside outside of peer-reviewed HCI publications. We found their traces in less prominent genres of HCI research dissemination, e.g., in <Interactions> magazines~\cite{Nathan_Interactions2010_RwandaPolicy,Lazar-HCI-policy-Interacitons15-forum}, in non-archival policy white papers~\cite{Policybrief2023_GenAI_Jobs}, in the news~\cite{FacebookPolicy_CHI16}, and on Medium~\cite{consequence-scanning-Salesforce}.
Unfortunately, these publications also provided little detail about the HCI-policy collaboration contexts, processes, or outcomes.}

\movedto{Other collaboration efforts seemed entirely untraceable.
For example, we suspect HCI research has influenced Facebook's content moderation policies, because many HCI scholars worked on the topic as Facebook employees. However, such collaboration is difficult to verify, much less to study rigorously.}

\subsection{Community Inputs and Discussions}
To collect more data on HCI-policy collaboration processes and breakdowns, we held a workshop on this topic at the 2023 CHI conference. With an open call for participation, the workshop brought together $57$ HCI researchers and practitioners from the Americas, Asia, Australia, and Europe. 
The workshop received and published more than $60$ papers detailing the participants' specific research projects, all at some intersection of computational systems, people, and policy.
They cover an overwhelming breath of topics, e.g., social media’s role in human trafficking, surveillance of migrant workers, unions negotiating tech policy for workers, combined use of HCI and policy in tenant protection, policy issues throughout the supply chain of generative AI, AI in e-governments, challenges in applying GDPR to user interface design, and many more. 
These publications became an additional source of data for our analysis.

\edits{
\subsection{Data Limitations}
One limitation of our data is geographical. While workshop participants represent many regions globally, all authors of this paper are based in the U.S. and Europe, as are most workshop publication authors. All except one workshop author are based in democracies. Recognizing this geographical bias, we strongly encourage researchers from other parts of the world to help critique this work and share their perspectives on HCI-policy collaboration. 
}

\subsection{Data Synthesis}
\edits{
We synthesized a new problem-solution frame based on this relatively diverse set of empirical reports. As we will demonstrate in the remainder of the paper, this frame can (1) help explain the observed disparity in HCI-policy collaboration and (2) reveal new avenues of opportunity in addressing the disparity. It is (3) flexible, allowing HCI researchers and practitioners to derive specific actions according to the respective context of their policy engagement. Finally, even the authors lack consensus on whether this problem-solution frame is too progressive and controversial, or too obvious that is what HCI needs to do. In this sense, this frame can be (4) an effective seed for broader HCI-community-wide discussion.

This frame results from the authors’ year-long discussion within our small team, with inputs from wider our respective communities. Appendix \ref{appendix:analysisMethod} describes our deliberation process in more detail.
}

\section[Framing Policy as Integral to HCI’s Intellectual Concerns]{Policy as Integral to \\HCI’s Intellectual Concerns}
\label{chapter:finding}

\subsection{A Vision of the Future}

We envision a future in which HCI sees policy as integral to its intellectual pursuits, placing system-people-policy interaction not at the boundaries of HCI and policy, but nearer the center of HCI research, practice, and education.
This vision differs from prior framings in three important ways.

\begin{enumerate}[leftmargin=*]
    \item HCI will deliberately dissolve the boundary between its policy- and system-human-interaction-focused work. Instead, a wealth of hybrid methods and hybrid knowledge contributions will bridge and harmonize HCI's system, people, and policy expertise. 
    \item Each HCI project will choose to integrate system, people, and policy considerations to varying degrees and in diverse modes. For example, some projects will advance HCI's knowledge of system-human interaction, while providing policy implications. Some projects will leverage human-system interaction expertise and advance HCI's knowledge of policy design. Other projects will generate new knowledge on system-human-policy interaction, by bridging the three areas of expertise. In this future, peer-reviewed HCI research venues will critically assess and accept these diverse knowledge contributions. HCI education and practitioner methods will embrace these varied methods.
    \item HCI communities will actively coordinate these varied methods and diverse knowledge contributions, maximizing HCI's collective impact on real-world policy outcomes. 
\end{enumerate}

\edits{
\subsection{A New Problem Frame}



There are three key differences between this vision of the future and the current landscape of HCI-policy collaboration. These differences offer us new insights into why HCI-policy collaboration efforts have had disparate results.
}

\textit{\textbf{1) Lack of recognition for the \emph{inherent} tensions between HCI and policy methods, leading to a lack of recognition for trans-disciplinary methods that address the tensions.~}}
Peer-reviewed HCI research venues, practitioner-facing methods, and HCI textbooks did not always give due recognition to unconventional methods and knowledge that bridged HCI-policy expertise.
For example, while many peer-reviewed CHI and CSCW papers feature policy recommendations derived from PD workshops (a well-established HCI method)~\cite{stickdorn2012service,forlizzi2018moving}, none featured policy proposals derived from researchers' own synthesis (for example, by synthesizing policymakers' evidentiary needs and HCI's prior empirical work)~\cite{OccupyCHI_USpolicy_CHI12panel}. This may be because HCI has not yet acknowledged synthesis as an established HCI research method, or policy design as an HCI knowledge contribution. Consequently, methods that can bridge HCI-policy evidentiary gap remain largely visible in prominent HCI venues.

This lack of recognition can contribute to the disparity in HCI-policy collaboration.
It can disincentivize early-career researchers from participating.
It also makes HCI-policy knowledge less accessible, adding to why socializing with policymakers \textit{appears} to be the only way to gain such knowledge~\cite{lazar2017_gap_year,OccupyCHI_USpolicy_CHI12panel,HCI-policy-boundary-CHI19}.
Both effects privilege established HCI experts (who are more likely to prioritize long-term policy impact and relationship-building over near-future publications). Both privilege issues aligned with policymakers' existing interests over precautionary debates.

\textit{\textbf{2)~Challenges in deliberating and curating novel HCI-policy methods and knowledge contributions.~}}
The limited visibility of novel HCI-policy methods and contributions not only hinders methodological innovation. It also inhibits broader HCI communities from scrutinizing or deliberating upon these emergent methods and findings. For example, while many Speculative Design and PD projects have discussed their implications for policy, few have asked: Are ``implications for policy" a necessary or good measure of the quality of empirical work that studies technological harm? What makes an implication for policy more meaningful or useful than others? 

The lack of deliberation on emerging HCI-policy methods means fewer proven method choices for HCI professionals attempting to integrate policy into HCI work. Little publicly available guidance exist on how to write better ``Implications for Policy." HCI practitioners' toolbox contains few methods for designing system-people-policy interactions.
This lack of public knowledge yet again disadvantages early career HCI professionals.

\textit{\textbf{3)~Insufficient community-wide coordination for collective impact.~}}
Efforts to coordinate HCI's diverse modes of policy engagement are scarce in HCI literature. Considering that policies underpin every computational system and every human, when is policy design \textit{not} the best approach to improve a system's human impact? Knowing that momentum for policy change can arise abruptly, how can HCI quickly mobilize its diverse HCI-policy work to seize the opportunity?
Answers to these meta-questions can amplify HCI's collective influence on policy outcomes, yet are noticeably absent in today's HCI research discourse.
    
With little community-wide coordination and support, HCI researchers and practitioners relied on their respective efforts for policy impact, exacerbating disparities in HCI-policy collaboration and weakening the collective impact of HCI on policy.

\subsection{A New Solution Frame}
\label{chapter:future}
These new problem frames reveal new solution frames; new strategies for addressing the disparity in HCI-policy collaborations.

\begin{enumerate}[leftmargin=*]
    \item Explicate underlying tensions between HCI and policy methods;
    \item Give due recognition to the trans-disciplinary methods that effectively address these tensions. Assess and accept their knowledge contributions to peer-reviewed HCI research venues, practitioners' toolboxes, and HCI textbooks; 
    \item Foster a mosaic of trans-disciplinary methods and knowledge contributions that blend policy and HCI's futuring, empirical, and design expertise to varying degrees and in diverse ways. Each individual HCI project can choose among them;
    \item \movedto{Coordinate HCI's diverse policy engagements and maximize HCI's collective impact on policy outcomes.}
\end{enumerate}

The remainder of the paper demonstrates the usefulness of this new problem-solution frame. We illustrate how even seeing existing literature through this new frame can reveal tangible, new opportunities for individual HCI projects (Chapter \S\ref{chapter:finding_usefulness_individual}) and for HCI communities as a whole~(Chapter \S\ref{chapter:finding_usefulness_collective}). 

Do HCI communities endorse this re-imagined HCI-policy relationship? Are these strategies indeed effective in fostering HCI-policy collaboration and improving computation's societal impact? Addressing these questions takes time and requires community-wide HCI efforts and debates. The opportunities outlined initiate such efforts and debates.



\edits{
\section[Opportunities for Individual HCI Projects]{Opportunities for \\Individual HCI Projects}\label{chapter:finding_usefulness_individual}
}

Seeing policy as integral to HCI's intellectual endeavors---be it revealing the impact of existing technologies (“\textit{empirical work}”), improving this impact by creating new technologies (“\textit{design work}”), or speculating the societal impact of emerging technologies (“\textit{futuring work}”)---can enhance these endeavors.

\movedfrom{We describe one emergent HCI-policy hybrid research method that some may consider too progressive, or too tilted towards policy methods and not enough towards HCI. By laying out a range of possible changes, we invite HCI communities to discuss and debate: Where do they think an ideal HCI-policy partnership should stand on the spectrum of HCI's and policy's current intellectual concerns?}

 

\subsection{Integrating Policy into Empirical Work}

\paragraph{HCI-Policy Synergies and Tensions.~}
Seeing policy as integral to HCI's pursuit of human understanding helps us see the connections and tensions between them.
Both policy and HCI communities want to understand people's values, behaviors, societal processes, and interactions with and experiences of emergent technologies.
\edits{HCI and policy share empirical methods such as participatory workshops~\cite{young2019toward,HCITacticsForPolicyCHI21Whitney}.
In many cases, they are already collaborating, e.g., in understanding misinformation on social media \cite{haimson2021disproportionate} and promoting gig worker welfare \cite{lee2015working,Savage2023_forecasting}.
}

\edits{Nevertheless, tensions may arise between HCI and policy actors' empirical work, because of their different scopes.}
\movedto{HCI's empirical findings illustrate \textit{situated} interactions among a specific combination of systems, people, policies, and contexts.} 
They mean to inform system designs that meet the needs of specific stakeholders in particular contexts. 
In contrast, policy design influences a broader set of technologies, people, and interactions, thus requiring larger-scale empirical evidence. This difference in scope has led to some frustrations in the HCI community that policymakers seek difficult or impossible empirical evidence, such as large-scale evidence of an emergent technology’s societal harm~\cite{HCI-policy-boundary-CHI19}.

\paragraph{Low-Hanging Fruits in Addressing HCI-Policy Tensions.~}
By explicating the tensions between HCI and policy's human understanding work, we can identify first steps in alleviating these tensions.
\edits{For example, HCI empirical work can communicate the generalizability of its policy implications more explicitly.} 
Is an observed technological harm specific to this particular technology or population? 
Is it the concern of global, national, local, or sectoral regulations?
Does it require an update to the spirit, the text, or the implementation of the law? 
Answers to these questions are within the reach of existing HCI empirical methods, yet can address policymakers' needs more directly.

HCI can also expand its empirical methods, observing how a specific policy helps shape various human-computer interactions, thereby increasing the likelihood of deriving generalizable policy implications. 
What limitations of current AI regulations does generative AI expose? Is a new privacy policy in conflict with the social norms of particular populations, such that they use technologies to circumvent it?

Expanding HCI's use of metaphors to communicate policy implications represents another tangible opportunity.
HCI communities have long used metaphors when communicating a technology's affordances and human impact.
Less discussed is that metaphors often emphasize different social values and carry rich policy implications (e.g., Are generative AI models more like Internet search engines (therefore their data contributors enjoy similar rights) or black boxes? Is attaching a GPS surveillance device to a car more like following a car on public roads (therefore does not require a court-ordered warrant) or more like trespassing on one’s private property?~\cite{US_v_Jones}). 
\movedto{In tech policy discussions, various interest groups often debate the choice of metaphors, because this choice anchors how policymakers understand a technology's affordances and decide whether it needs new policy and legal frameworks~\cite{wong2015wireless}. }
There is a ready opportunity for HCI to improve its communication with policymakers via metaphors. It can help link HCI's contextual empirical insights to broader policy contexts, \movedto{using a language that they are familiar with.} 
\movedfrom{How does the impact of generative AI on people differ from previous waves of AI advances? Are there legal precedents and historical parallels one can draw?}

\movedto{
HCI's empirical work might also take advantage of the fact that law and policy differ across geographic scales and jurisdictions, linking local observations with broader policy implications.
Take, for instance, the data breach notification laws in the U.S. Rather than a single national law governing data breaches, this set of patchwork rules was created and passed on a state-by-state basis over sixteen years (2002-2018) \cite{securityBreach_notification_laws_2022}. 
This offers a distinctive opportunity for HCI empirical work. On one hand, HCI can conduct natural experiments, comparatively analyzing how technologies and different laws and their implementations play out differently. On the other hand, because companies and stakeholders have the incentive to comply with the strictest local policies that impact their technologies, HCI can leverage policy elsewhere to promote change in their locale of interest.}

\paragraph{Emergent HCI-Policy Trans-Disciplinary Methods.~}
Seeing policy as integral to HCI also allows us to imagine not-yet-existent empirical methods to study system-people-policy interaction. 

For example, we see an opportunity for empirical research on successful technology and policy designs. 
Good designs of technologies, social practices, and policies fade into the background of everyday life, making them more difficult to observe (e.g., via stakeholder interviews). Presently, these success stories can offer valuable lessons, yet are almost entirely absent from HCI literature.
A key aim here is to make what is generally invisible to lay people visible, to allow technology designers, policymakers, users, and other stakeholders to reflect on the kinds of system-people-policy interaction design they would find meaningful and actionable.


Next, if HCI is ready to embrace passive, observational empirical work on system-people-policy interaction, are we ready to accept research that creates probes to study tech-policy interactions?
\begin{quote}
\textit{Research through Litigation.~}

\citet{CHI23WS:ResearchThroughLitigation} is a computer scientist with seven years of experience bringing legal cases to Courts and Tribunals. He does so for the purpose of (1) understanding how the law and legal system operates on the ground and, secondarily, when possible, (2) changing the law. He named this approach “Research through Litigation.”

The idea underlying this approach is that the law is a set of ill-defined rules that variably apply to changing real-world circumstances. We cannot understand how a law works simply by reading the law on paper. Further, good laws often fail because of a lack of institutional capacity to implement them.

To understand this gap between law on paper and law on the ground, Kirkham carefully picked legal cases that are good “\textit{test cases.}” The legal process allows him to argue for his case, observe people's reactions, and, in the process, generate voluminous documents (e.g., detailed correspondence within tech companies, expert testimonies) unlikely to be obtained through other means. He then analyzes this process using autoethnography and document analysis methods. This approach has surfaced “\textit{some serious (and somewhat surreal) concerns with the operation of the justice system}” and indeed changed the law on several occasions~\cite{CHI23WS:ResearchThroughLitigation}.
\end{quote}

People's experience of technology is highly context-dependent. HCI empirical researchers routinely use artifacts (e.g., technology probes, cultural probes) to give users and stakeholders a taste of the future. In so doing, researchers gain better insights into how people might interact with future technologies and derive more informative design implications.  

People's experience of policies is also highly context-dependent. The proposal of Research through Litigation highlights this complexity that most prior HCI-policy research neglects. In a sense, the legal cases Kirkham created are analogous to technology probes. They mean to reveal how law works on the ground, and on occasion, even improve the law. 
In this light, does this seemingly radical method merit more consideration?
Can HCI effectively simulate how courts operate, as a way to assess the effects of new tech and new policy on the ground?

\begin{figure*}[b]

\vspace{0.2cm}
    \includegraphics[width=0.7\textwidth]{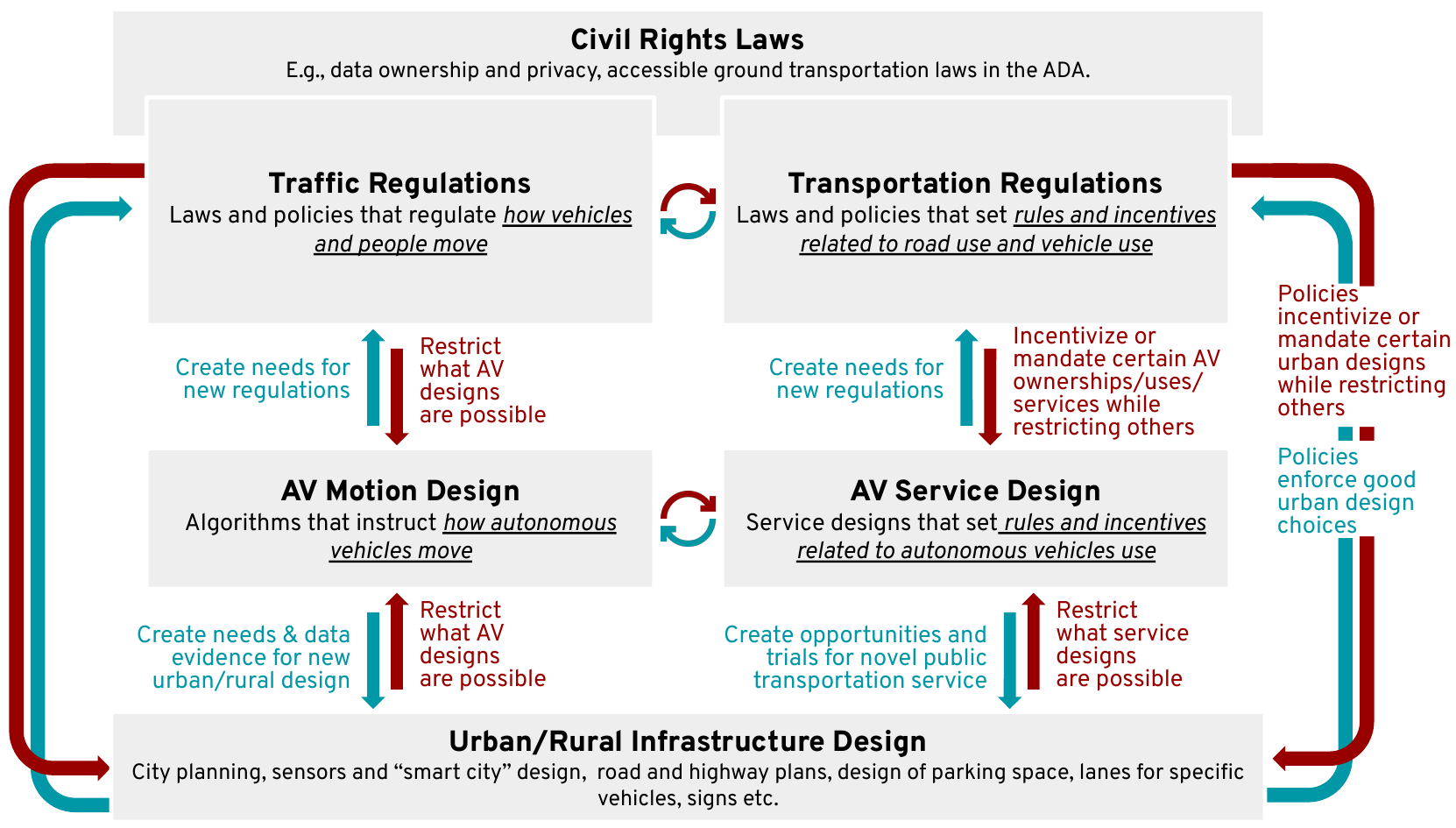}

      \captionsetup{width=0.8\textwidth}
  \caption{The many ways autonomous vehicle (AV) behavior design, city design, and law and policy interact in the U.S. legal context. They enable (green arrows) and constrain (red arrows) each other~\cite{CHI23WS:Sandhaus_AVCityPolicy}. }
  \Description{A diagram mapping the many ways autonomous vehicle motion design, city design, and law and policy influence each other. They enable (green arrows) and constrain (red arrows) each other.}
\label{fig:interactions}
\end{figure*}

\subsection{Integrating Policy into Design Work}
\edits{
\paragraph{HCI-Policy Synergies and Tensions.~}
As mentioned earlier (\S\ref{chapter:def_understandDesignPolicy}), HCI and policy design share many synergies: Both seek to improve technologies' human impact. Both must thoughtfully shape their design processes, because design processes anchor design outcomes. Both must navigate the tensions between design goals and technical feasibility constraints~\cite{feasbility_gap,hayes2002policy_change_limits_incrementalism}.
HCI and policy designers already share formal methods such as participatory design~\cite{delgado2023participatory,HCITacticsForPolicyCHI21Whitney}.

Seeing policy as integral to HCI's design inquiries illuminates additional connections between them.
For example, we realize that both HCI and policy use service design methods~\cite{junginger-book}.
}\movedto{
Although service designers do not always mention the word policy, they routinely design computational systems, organizational workflows, and related corporate or platform policies in tandem~\cite{stickdorn2012service,forlizzi2018moving}. When service designers specify a \textit{freemium} model for a new app, they specify policies around who can access what quality of service at what price, a simple micro-economic policy. 
They shape user experience by innovating system-human-policy interactions. 
}

Other forms of HCI-policy design collaboration are also emerging. For example:

\begin{itemize}[leftmargin=*]
    \item \textit{Designing public service/policy.~}The boundary between designing computational systems for government agencies and shaping public policy is porous. Therefore, many HCI design projects in the public sector involve policy design. For example, researchers redesigned the U.S. Postal Service (USPS) manual and, in doing so, reshaped policies around how small businesses can access postal services, alleviating their legal burdens~\cite{junginger-book-USPS}.
    \item  \textit{Designing both computational systems and policies based on a shared strong concept.~}\movedto{For example, FAccT publications on what “fairness”~\cite{solon2017fairness} or “representativeness”~\cite{chasalow2021representativeness} means catalyzed synergistic designs of AI systems and their regulations.}
    \item \textit{Designing policy evaluation metrics.~}\movedto{For example, \citet{CHI23WS:Jin_FAccT_Criminal} collaborated with legal professionals to define evaluation criteria for an algorithm in the U.S. criminal legal system. This design bridged HCI and policy experts' notions of a good algorithm and a good legal system.}
    
    \item \textit{Designing legally mandated tech design processes.~}Some HCI researchers promoted HCI's human-centered design processes to become legally required or recommended technology design processes~\cite{OccupyCHI_USpolicy_CHI12panel}.

\end{itemize}



Tensions can also arise between technology and policy design, for example, due to documentation differences. HCI and policy communities differ significantly in how they document, disseminate, and attribute their designs. One communicates design ideas via computational artifacts, demos, and pictorials. The other uses policy white papers and law reviews. As a result, social connections (e.g., an HCI designer taking a gap year to work at government agencies) can seem like the only route to simultaneously engage in technology and policy design. 

Secondly, coordinating HCI and policy design processes is challenging, because of their multiple and disparate temporal patterns.
\movedto{Innovations in HCI system design can occur rapidly, but may also progress slowly, for example, when developing large computational infrastructures. Policy change can culminate over decades, but can also suddenly accelerate, for example, when public interests surge (e.g., on regulating generative AI~\cite{chatGPT_quickRegulation}) or certain public events occur (e.g., the outbreak of COVID-19~\cite{Congress_COVID_contact_tracing}).}

Lastly but crucially, power differences complicate synergistic HCI-policy designs. Corporate and public policies can shape computational system designs and user interactions with their authority and power. The reverse is less true. 

\paragraph{Low-Hanging Fruits in Addressing HCI-Policy Tensions.~}
By explicating the tensions between HCI and policy design,  we can identify practical solutions to address them. For example, improving documentation of HCI-policy design collaborations. Prominent HCI research venues can contribute to this by accepting coordinative tech and policy designs into their proceedings. 

Other near-future solutions can further support and popularize the emergent forms of HCI-policy design collaboration listed above. For example, we see an opportunity to extend the practice of designing legally mandated tech design processes into the context of privacy by design.
\deletes{For instance, GDPR's “data protection by design and by default” requirement~\cite{jasmontaite2018data}; the “comprehensive privacy and security programs” instituted by the FTC \cite{ftcgov_privacy2020}; and the “Risk Management Frameworks for data privacy, security, and AI" created and recommended by the US National Institute for Standards and Technology.}
While policy documents broadly call for "privacy by design", it is not always clear \textit{how} that should be done. Consequently, legal assessment and risk management practices ended up operationalizing these design tasks~\cite{wong2019bringing}.
HCI researchers have already created many design methods to address privacy, and can more explicitly connect their work with the goals of privacy by design as articulated in policy documents.
More ambitiously, HCI researchers can actively promote HCI design methods (e.g., user-centered design, value-sensitive design) as equally essential approaches, on par with legal and risk management strategies in privacy by design.


HCI designers can strategically align their tech design goals with policymaking timelines.
Upon technical and/or social change, HCI's novel system design contributions can help fill in the temporary policy vacuums \cite{moor1985computer}, bridging the gap between what the law protects and promotes and what people believe the law ought to protect or promote \cite{bamberger2011privacy}.
In doing so, HCI's system designs can help protect people and circumstances that the law fails to protect~\cite{feasbility_gap}. 

\paragraph{Emergent HCI-Policy Trans-Disciplinary Methods.~}
Seeing policy as integral to HCI sets the stage for HCI and policy designers to collaboratively design technologies, social practices, and policies in tandem, creating an artful tech-policy-human interplay.
Service designers already do so to an extent, though typically within the confines of one company's technologies and policies~\cite{policy-generativity-Centivany-CSCW16,pillowfort-platform-design-and-policy-chi22}.
There is an open opportunity to expand the emergent practice of simultaneous tech-and-policy design to larger-scale design contexts.
\deletes{For example, to build a healthier social media environment, what new solutions might emerge if HCI and policy experts collaborate on prototyping new content moderation AIs, new platform policies, and new government regulations? To create livable future cities, what new solutions might emerge if HCI and policy experts concurrently and collaboratively autonomous vehicles, smart city infrastructure, and related traffic and road policies?}
\begin{quote}

\citet{CHI23WS:Sandhaus_AVCityPolicy} proposed iCAPS (Integrative Prototyping of City Environments, Autonomous Vehicle Behaviors, and Policies), a simulation-based prototyping platform.
Build upon the digital twin of a city, this platform simulates and visualizes how design changes to autonomous vehicles’ (AVs') driving behaviors, city design, and related policies might influence city design goals (e.g., road safety, neighborhood equity, pollution).
Via simultaneous AV-city-policy design, this platform promises to catalyze safer, more equitable, and more sustainable future cities~\cite{Gilbert-sociotechnical-specification}. 

This platform is also a boundary object.
AV, city, and policy design decisions belong to many different private corporations and government agencies.
\edits{Although proven beneficial~\cite{Riggs2019-lx}, coordinating these design decisions remains challenging in practice.}
\deletes{While AV research shows even small tweaks to traffic light and road sign design could instantaneously improve AV computer vision performance and increase safety~\cite{Riggs2019-lx}, AV designers simply are not in the position to adopt these designs.}
The iCAPS platform addresses this challenge by bringing AV, city, and policy designers together and moderating their design actions.

\end{quote}
This emergent method embodies the idea of designing technology and policy simultaneously, and proposes one concrete way of operationalizing it. 
It raises at least two sets of useful questions.

This method highlights the power struggles among designers when they design technology and its many related policies in tandem.
Systemic thinking is integral to HCI design expertise. It allows designers to grasp how various laws, policies, technologies, and contextual factors can collectively influence people's experiences~(Figure~\ref{fig:interactions}).
However, it would be naive to think that tech designers can re-design related laws and policies without multiple policy experts at the table. 
Must designers convene \textit{all} relevant policymakers to the same table, in order to design tech and the many policies it involves in tandem?
If that is not feasible, which laws and policies should HCI designers prioritize? 

This method also underscores the challenges of evaluating system-people-policy interaction design.
While HCI design typically iterates and learns from failures, a failed public policy can have irreversible consequences~\cite{wicked_problems_rittel1973}.
The proposed prototyping platform addresses this challenge through simulation, computationally predicting societal outcomes for each AV-city-policy design. This contrasts with HCI's traditional approach of evaluating designs with real stakeholders. Do HCI communities support this shift?
%
\subsection{Integrating Policy into HCI Futuring Work}

\paragraph{HCI-Policy Synergies and Tensions.~}
Both policy and HCI communities want to foresee how emergent technological capabilities might interact with people and societies, cause disruptions, and create needs for new computational system designs, social practices, and/or policies. 
\deletes{Typically, this work focuses on technological harms.}
To do so, the field of HCI has methods such as speculative design and consequence scanning~\cite{consequence-scanning-Salesforce,UK_open_policy_toolkit}. Policy actors have methods such as forecasting~\cite{Savage2023_forecasting}
and the Precautionary Principle~\cite{kriebel2001precautionary}.

Yet at least two differences between HCI's and policy's approaches can hinder their collaboration.
Prior research often criticized HCI's policy engagements being too reactive~\cite{policyknot-CSCW14}. 
These hindrances might explain why.

First is their different views of risks.  
HCI's speculative designs indicate qualitative risks, often without indicating the level of urgency. Their goal is to provoke discussion and providing cautionary tales for new technologies' adoption and (mis)use.
Risk and urgency levels, however, are essential for policy actors' futuring work: Is a new technology's societal harm catastrophic but unlikely (like nuclear war), slowly unfolding but inevitable (like climate change), or so catastrophic and inevitable that policy should forbid it outside of research labs if not entirely (like genetic editing of deadly diseases)? The answer to this question is crucial for policy-making~\cite{jasanoff2014serviceable}, yet fall outside of existing HCI futuring methods.

Second is their different views of design. As we have hinted at earlier in the paper, while speculative design is a process of open-ended knowledge inquiry~\cite{wong2019bringing}, some legal scholarship sees design as merely the implementation of a given policy requirement.
Many policymakers engaged designers only after they had identified a policy goal (e.g., ensure citizens' right to privacy) and had translated it into system requirements (e.g., providing individual control over personal information by increasing consumer notice and choice \cite{gellman2022fair}). They engaged HCI designers only to implement these requirements (e.g., design the standardized cookie opt-out menus~\cite{Tkacik_2020}).

\paragraph{Low-Hanging Fruits in Addressing HCI-Policy Tensions.~}
By explicating the connections and tensions between HCI and policy's futuring methods, we start to identify easy ways to strengthen the connections and alleviate tensions.
\deletes{HCI's futuring work already suggests policy implications. We envision this work to explicate these connections, even when using existing futuring methods.}

\edits{For example, HCI's speculative design work can articulate its <Implications for Policy> more explicitly and concretely.}
\deletes{For example, after identifying emergent socio-technical issues, they can summarize and translate into concrete policy problems:}
What emergent technologies (e.g., facial recognition AI) might require new methods to meet old policy goals? In what ways may current laws and policies fall short in addressing a problem? What values that people hold do not fit into existing legal frameworks (e.g., changing privacy norms)? How might policymakers define a new policy problem space (e.g.,~“dark patterns”~\cite{gray_darkpattern_CHI18}), based on the emergent socio-technical issues this work has identified? Are new policy initiatives necessary~\cite{lindley2017anticipating}? 
\deletes{As trivial as it sounds, this translational work is a pragmatic solution to HCI work’ and policymakers’ methodological gaps that previously hindered HCI’s policy impact~\cite{HCI-policy-boundary-CHI19}. }

Explicating policy implications of HCI's speculative design work can potentially increase its policy impact.
In addition, this approach can showcase the value of speculative design throughout the policymaking process, gradually correcting misconceptions about HCI's design expertise among some policy actors.
Imagine, if HCI researchers had defined the new social issues and policy needs that chatGPT would entail, within the four years after its neural architecture was first published in academia and before its public release. These insights might just tackle the challenge of ``\textit{bringing policymakers to the table}''~\cite{HCI-policy-boundary-CHI19}.


\paragraph{Emergent HCI-Policy Trans-Disciplinary Methods.~}
Seeing policy as integral to HCI's futuring work also gives us license to imagine entirely new types of futuring research.
For example, \deletes{imagine if CHI will accept forecasts of emergent technologies' societal risks---based on rigorous analyses of the technology and trend, an analysis of current human practices and how they are likely to be impacted, and analysis of how they will interact with existing laws and reveal policy gaps---as a new type of HCI research. 
In addition, we envision }speculative design practice might embrace speculations around law and policy. It can speculate about the social and technical implications of an emergent policy proposal. It can also envision new uses and forms for emerging technologies, new categories of law and policies, and new social practices. After all, these factors and the interactions among them simultaneously shape emergent technologies' societal impact.


\edits{More ambitiously, we see opportunities in exploring new technological risk forecasting methods that combine HCI's qualitative, provocation-oriented approaches with policy's semi-quantitative, action-oriented approaches. To help provoke our imaginations around these opportunities,}
\movedto{we describe one such method emergent from our workshop; a method that some may consider too progressive, or too tilted towards policy methods and not enough towards HCI.}
\deletes{By laying out a range of possible changes, we invite HCI communities to discuss and debate: Where do they think an ideal HCI-policy partnership should stand on the spectrum of HCI's and policy's current intellectual concerns?
Example New Method: AI Capability Forecasting.~Research that combines HCI and policy expertise for technology development and risk forecasting is already emerging at the periphery of HCI.}
Presenting this method as a provocation, we ask: Suppose we (HCI communities) are ready to embrace policy as an integral part of its futuring work. Are we also ready to embrace political science's view of rigor in forecasting?

\begin{quote}
\textit{AI Capability Forecasting:~}

\citet{CHI23WS:BCG_Forecasting}~proposed to quantitatively forecast AI capability growth, in order to change the reactionary stance HCI researchers, policymakers, and organizations often found themselves in when regulating AI. Such forecasts, and the process of generating and debating them, can help them build mid- and long-term strategies for addressing AI's societal impact.
The quantitative forecasting method has three steps:

\begin{enumerate}[leftmargin=*]
    \item Define a technology capability prediction problem with clear evaluation criteria;
    \item Invite many researchers and forecasters with a variety of relevant expertise to submit their quantitative predictions, along with documentations of their methods, assumptions, and uncertainty measures;
    \item Aggregate the forecasts by examining their degree of consensus. As a secondary step, some researchers bring forecasters together, enabling them to challenge each others' assumptions, sharpen their methods, and collaboratively converge on the most likely predictions.
\end{enumerate}
\end{quote}

This method challenges HCI's assumption that speculations of technology harms are only qualitative. 
To inform policy actions, analysis of how likely or how soon the harm will become a reality is critical. 
\movedfrom{Is a new technology's societal harm catastrophic but unlikely (like nuclear war), slowly unfolding but inevitable (like climate change), or so catastrophic and inevitable that policy should forbid it outside of research labs if not entirely (like genetic editing of deadly diseases)? The answer to this question is crucial for policy-making, especially precautionary policymaking \cite{jasanoff2014serviceable,kriebel2001precautionary}.} 
In order to impact policy, is HCI's speculative design work open to alternative criteria of rigor in its speculations?

Throughout Chapter~\ref{chapter:finding_usefulness_individual}, we illustrated that seeing policy as integral to HCI's intellectual endeavors can enhance these endeavors.
Examining existing HCI literature (including ones previously considered at the fringes of HCI) through this new lens, we can begin to see inherent tensions between HCI and policy methods, identify pragmatic approaches to alleviate these tensions, and deliberate on emergent methods that address these tensions in thought-provoking ways.
Going forward, we encourage HCI communities to together foster a mosaic of distinctively HCI methods and knowledge contributions that blend system, human, and policy expertise to various degrees and in diverse ways, creating robust, grass-root connections between the two fields. Chapter~\ref{chapter:finding_usefulness_individual} outlines an initial draft of this new landscape.

\section{Opportunities in Enhancing HCI's Collective Impact on Policy} \label{chapter:finding_usefulness_collective}

Now, we turn our attention to HCI-policy collaboration at a community level.
We previously proposed that, by seeing policy as integral to their intellectual concerns, HCI communities will coordinate their diverse policy engagements, thereby amplifying HCI's collective impact on policy outcomes.
In this chapter, we outline four emergent opportunities for such coordination by synthesising existing HCI and policy literature.
\deletes{We envision a community-wide HCI-policy collaboration that differs from the status quo in four critical ways:}



\subsection{When (Not) to Policy}
\movedfrom{Policy interventions are not a cure-all.}
\deletes{Coordinating tech and policy design is also not always the right thing to do.}
\movedfrom{As obvious as this may sound, this is missing perspective from the current HCI literature.}
\deletes{We envision a future where HCI reflectively chooses among policy and other means to address technologies' societal impacts, based on a principled understanding of policy's strengths and limitations.}
\edits{We envision a future where HCI communities strategically choose when and when not to use policy to address computation's societal issues. A crucial step toward realizing this vision is establishing a principled understanding of policy's affordances and limitations, compared with other HCI's tools. There is an under-explored area in HCI research and a clear opportunity for future work.}

On the one hand, HCI can further exploit the affordances of policy.
For example, current HCI-policy work focused overwhelmingly on regulating technology harms. Less discussed is how policy can also enable new actions. For example, copyright law attempts to balance protecting creators from intellectual property harms, while also encouraging new innovations for social good through fair use. How might HCI leverage policy to promote the positive impact of technologies?

On the other hand, \movedto{policy interventions are not a cure-all, yet their limitations are overlooked in current HCI literature.}
They can be prone to their own forms of inequities and bias~\cite{DotheRightThing_CACM2023}. %
Moreover, law and policy solutions promote incremental changes within an existing legal and institutional structure. They typically do not “change the system,” particularly not overnight~\cite{bork2019evolutionary}. Depending on the degree and kind of social change HCI hopes to create, sometimes, other approaches (e.g., community organizing) might be more effective than policy interventions.

\edits{Explicating the limitations of policy can also reveal new opportunities for technology design.}
For example, HCI researchers often criticize that capitalist goals drive many technology product designs and cause ethical harm. They then seek remedies from policy interventions.
However, law and policy \deletes{can be prone to their own forms of inequities and bias.
Sometimes, they }can be just as enwrapped in promoting or serving institutions of capitalism~\cite{regulatory_capture2006}.
\deletes{Vulnerable populations can find themselves in precarious situations where both tech and policy coordinately put them at a disadvantage \cite{CHI23WS:Tran_adversarial}.}
In order to improve technologies' societal impact, HCI sometimes needs to consider designing technologies that help the disenfranchised populations fight unjust policies~\cite{CHI23WS:Tran_adversarial}.

\subsection{Which Mode of Policy Engagement \& When}

\deletes{The previous chapter has laid out many “\textit{strong ties}” linking policy and HCI's intellectual pursuits, as well as many “\textit{weak ties}” where existing HCI methods can offer policy insights.}
\deletes{Below, we offer our perspectives on the modes of collaboration suitable for various technologies. This is one of the areas where disagreements remain among the authors. We encourage the HCI communities to critique the ideas below and share their answers.}

\edits{
We envision a future where HCI communities will wisely choose the extents and modes of policy engagement, based on the specific technology and social issue at hand.
A crucial step toward realizing this vision is to develop a principled understanding of \textit{how} the most productive approach to policy engagement differs according to context.
For example, when is appending a thoughtful <Implications for Policy> to a classic HCI empirical study already sufficient? In what contexts is establishing deep, ongoing partnerships with policymakers not just beneficial, but necessary? 
}

The history of HCI---more specifically, the three waves of HCI~\cite{bodker2015third}---can serve as an initial scaffolding for this discussion.
This history seems to suggest that as a computing system becomes more deeply and broadly integrated into societal processes, addressing its impact requires deeper collaboration between HCI and policy.

\deletes{Using the four waves of HCI as a scaffold~\cite{bodker2015third}, we can see the history of HCI echoes this argument:}

\begin{itemize}[leftmargin=*]
    \item The first wave of HCI focuses on making computers more efficient and functional for computer experts. These efforts involved minimal policy considerations.
    \item The second wave of HCI focuses on making personal computers easier to use and more accessible to everyday users. These efforts catalyzed the ISO usability standards and accessibility laws, which in turn enhanced later HCI work. Here, HCI and policy work operated asynchronously. Appending a thoughtful <Implications for Policy> to an HCI empirical study on usability or accessibility could go a long way.
    \item The third wave of HCI focused on weaving computing into daily life through smartphones, tablets, and wearable devices. Around this time, HCI privacy scholars like Lorrie Cranor drafted national privacy policies at the FTC and designed standardized cookie opt-out menus~\cite{OccupyCHI_USpolicy_CHI12panel}. State laws later recommended these designs~\cite{Tkacik_2020}. Here, while HCI and policy work remained somewhat asynchronous, technology and policy design processes started to overlap. 

\end{itemize}

Do the three waves of HCI effectively stratify technologies' needs for HCI-policy collaborative work?
Is there a need for additional or more fine-grained categories?
These questions provide fertile ground for future research.

For example, the societal impact of interactive cyber-infrastructure appears to demand particularly close HCI-policy partnerships~\cite{Jackson2007understanding_infra}, and therefore might merit its own category.
Interactive cyber-infrastructure refers to interactive systems that not only host, \textit{but govern} various interactions on a societal-level scale\footnotemark. 
For instance, social media platforms whose algorithms govern online discourse, gig work platforms whose algorithms govern work assignments, pervasive sensing networks that govern how smart cities operate, AI Foundation Models (FMs) such as chatGPT that govern downstream AI innovations, and virtual reality (VR) “\textit{universes}” where governments offer public services \cite{AccentureVRgov_2023}. 
Understanding and improving the societal impact of cyber-infrastructure necessitate close HCI-policy partnerships, for three reasons. 

\begin{itemize}[leftmargin=*]
    \item Because of the immense scale of these systems, HCI's conventional empirical methods may struggle to trace or prove their societal impact. Law and policy methods (e.g., Research through Litigation) might help;
    \item Because of the immense power the cyber-infrastructure owners wield (think the power of OpenAI), HCI designers heavily rely on policies that mandate system transparency and/or access, in order to do human-centered design work. Designing one more socially beneficial GPT prompt is simply not as impactful as devising a way of reducing the creation of harmful prompts. The latter necessitates policy interventions.
    \item Because these systems are infrastructures, they shape how HCI researchers and practitioners work. \textit{Should CHI keep publishing novel GPT applications, even though they \emph{might} be soon regulated out of existence?}~Questions like this highlight the necessity of HCI-policy considerations from the outset. 
\end{itemize}

\footnotetext{To be clear, here we use the term "\textit{interactive cyber-infrastructure}" to denote a distinct category of computational systems, rather than as an analytical lens on the sociotechnical systems that shape society. 
We use the word ``\textit{infrastructure}'' differently than in Science and Technology Studies (STS) and cultural studies (e.g., ~\cite{Jackson2007understanding_infra,star1994steps,InfraProblemInHCI_CHI10}). }

\subsection{How to Maximize HCI's Collective Impact}\label{finding_coordination}
We envision a future where HCI can assemble the different subsets of its work to strategically influence different stages of policy-making, and to inform different policy actors~\cite{junginger-book,kingdon1984windows_of_opportunity}.
When sudden momentum for policy change appears, HCI communities can quickly assemble their diverse relevant work to seize the opportunity.

\movedfrom{HCI literature commonly suggests that the pace of technology development is fast while policy development moves slow~\cite{OccupyCHI_USpolicy_CHI12panel}.
Yet in fact, law and policy, technology development, and HCI advances all have multiple temporal patterns; They all sometimes move slowly and at other times move quickly~\cite{steinhardt2014reconciling,steinhardt2015anticipation,messeri2015greatest}. 
Just like research can move slowly, such as when developing large computational infrastructures and establishing new collaborations with stakeholders \cite{steinhardt2014reconciling,le2015strangers}, similarly, law and policy processes can have multiple temporal patterns, which can change over time.
By considering these different temporal patterns in both policy and HCI, we envision HCI can more effectively make a policy impact by both engaging with policy in a long-term, ongoing manner, and engaging in strategic ways when there is sudden momentum for law and policy change.}

\paragraph{Preparing the ground for policy change.~}
Even policy changes that seem to occur “suddenly” have longer histories of community discussion and debate. This period is an opportune time for cultivating relationships with policy actors, engaging in policy discussions that occur in the background of everyday life, and incubating HCI empirical and design work. 
\deletes{For example, the EU’s General Data Protection Regulation (GDPR) was adopted in 2016 and became effective in 2018, and as a sudden massive change, leading many organizations to quickly change their practices in response, in order to comply with the new regulations. However, the GDPR did not just happen on its own accord, it was the culmination of over 20 years of discussion and implementation of pre-cursor data protection directives in Europe. Therefore, it is advantageous for HCI to engage in these ongoing discussions that occur in the background of everyday life.} 

HCI researchers and practitioners can become more actively engaged in everyday policy discussions, for example, by responding to regulatory agencies' requests for public comment~\cite{FederalRegister_rulemaking} and writing Op-Ed articles (and other forms of tech journalism). In these discussions, HCI professionals can share metaphors that help policymakers understand the affordance of emergent technologies, offer empirical evidence of technological harm as expert evidence for future policies, recommend human-centered design methods to be legally mandated or recommended, and propose value-sensitive metrics for evaluating future policies. All this work can prepare the ground for future policy change.

In addition, the potentially long “incubation” period of policy change is also opportune for long-term HCI research, such as conducting participatory design workshops and comparatively analyzing how different tech-and-policy designs play out in different locales.

\paragraph{Catalyzing a groundswell of public interest and policy demand.~}
Law and policy can change quickly in response to current events (e.g., COVID-19~\cite{Congress_COVID_contact_tracing}), technical developments (e.g., chatGPT~\cite{Brookings_EUAIAct_Enforcement2023}), and other changes. 
\movedfrom{For instance, after the start of the COVID-19 pandemic, governments and private organizations quickly changed their policies and practices to create contract tracing systems. 
In 2023, just months after the public release of ChatGPT, questions about its potential effects on societies and industries created renewed interest among governments, schools, and companies to regulate AI systems~\cite{chatGPT_quickRegulation}. }
Yet that does not mean HCI has to wait passively for a pandemic to see windows of opportunity for policy change open up.


HCI's speculative, provocative, and participatory design work can catalyze a groundswell of public opinion around HCI issues, creating a new policy demand. Critical HCI is already somewhat successful in this regard~\cite{HCITacticsForPolicyCHI21Whitney}. Additional opportunities exist for HCI to engage interests groups, corporations, and policymakers more pointedly. The aim is to align the stars: When “\textit{a problem is recognized, a solution is available, and the political climate happens to be right}”, a window of opportunity for substantial policy change opens~\cite{kingdon1984windows_of_opportunity}. 

\paragraph{Seizing sudden momentum of policy change.~}
When a window of opportunity for policy change opens up, the heightened levels of attention are fleeting~\cite{kingdon1984windows_of_opportunity}.
HCI communities should be prepared to seize the political momentum. 

HCI researchers can prepare rapid policy responses by re-using their existing work, but adding a ``hook'' that connects it to the current events and political momentum. <Implications for Policy> sections of past HCI futuring work, literature reviews summarizing existing empirical evidence, and policy proposals tested in prior participatory workshops are all valuable, reusable resources here. 
These responses can take the form of op-eds, memos, policy white papers, public agency comments, and more. 


\subsection{How Might HCI Institutions Help}
Finally and importantly, we call for HCI institutions to nurture the ties between HCI and policy communities and to amplify HCI's collective voice in the policy world.

\paragraph{Including policy education in HCI; including HCI in policy education.~}
We envision a future where policy-making and policy implementation become a standard element of HCI education. This education can cultivate future HCI researchers and practitioners who can navigate layers of government agencies, navigate laws of intricately overlapping jurisdictions, connect with policymakers, and integrate policy expertise into their own work.
We also see an opportunity for HCI to become part of legal and political science education. Today, many leading law schools already offer curriculum on AI ethics, as do many HCI programs. Harnessing and expanding these existing connections incubate future HCI-policy collaborations of all sorts.

\paragraph{Celebrating small wins.}
From accumulating research evidence, to building relationships with policymakers, to acquiring policy expertise, HCI researchers and practitioners' journey to policy impact is arduous.
Moreover, such impacts are often invisible: Citing academic references or attributing individual researchers is not a convention in legal or public policy documents.
How can HCI institutions (such as CHI and CSCW) acknowledge and celebrate small wins in HCI researchers' policy engagement?

Consider: Can CHI proceedings accept insightful literature reviews that synthesize prior HCI works' <Implications for Policy> into actionable policy proposals? Can CHI offer a short paper track where researchers share their responses to regulatory agencies' calls for comments?
\edits{When public agencies or policymakers call for scientific evidence, they often require a response within days or weeks~\cite{OccupyCHI_USpolicy_CHI12panel}. In parallel to peer-reviewed publications, can HCI conferences create channels for faster-paced policy discourses and contributions?}
Such small recognitions and incentives have the potential to significantly boost policy engagements, particularly for early-career professionals in HCI.

 
\paragraph{Amplifying HCI's collective voice in the policy realm.~}
HCI institutions can help strengthen the collective ties between HCI and policy communities. \edits{For example, the registration costs for HCI conferences may be difficult for public servants to justify; are there ways to make the HCI community more accessible to these policy audiences?} Can CHI sets up booths for policymakers and public servants (just like it does for recruiters), helping HCI researchers connect with them?

HCI institutions can help amplify HCI communities' collective voices in the policy realm.
For example, HCI can offer social infrastructures that enable HCI professionals to collaborate on public comments, statements, and policy recommendations.
This approach can help more evenly distribute the policy engagement work across community members, and can amplify the collective voice of HCI communities.

\section{Closing Remarks}

\edits{The field of HCI has a history of self-improvement: identifying a problem, creating a new way of working, and then using evidence of success to argue for a change in the community.
Early computer systems caused chaos when they first entered workplaces, facing users who were not computer scientists. Early HCI researchers (before even calling themselves HCI) created a new way of designing technologies in response, that is, \textit{user}-centered design. 
When computing moved out of the workplace and into people's everyday lives, \textit{users} became \textit{consumers}, and they had choices. In response, HCI practitioners worked to improve the situated experience of technologies, and researchers dug in and unpacked the very concept of ``\textit{experience}''.

All this bodes well for the changes facing our community today. In current technology discourses, people are not only users and consumers. They are members of communities, citizens, and participants in lived democracy.
Technologies are not only systems. They are socio-technical platforms and cyber-infrastructure that govern societal processes.
People's ``\textit{experience}'' of technologies goes beyond efficiency or pleasure, but concerns the future of work, equity, sustainability, and more. Within such a milieu, we argue for a change in our community. 
We argue that HCI should re-examine the relationship among systems, people, and policy, and reflect on how HCI wants to position itself in this relationship. This paper provides an additional perspective to this unfolding dialogue.

}

\begin{acks}
Qian Yang's effort to help organize and participate in the workshop was partially supported by National Science Foundation under grants IIS-2212431, IIS-2313077, and Schmidt Futures' AI2050 Early Career Fellowship.
Richmond Y. Wong's effort was supported by a Georgia Tech Ivan Allen College Small Grant for Research (SGR).
Sabine Junginger's work on integrating policy and service design is supported by the Swiss National Science Foundation SINERGIA CRSII5 189955 VA-PEPR (Voice Assistants, People Experiences, Practices and Routines).
John Zimmerman's effort to help organize and participate in the workshop was partially supported by National Science Foundation under grants IIS-2007501 and IIS-2112633. Any opinions, findings, conclusions, or recommendations expressed in this material are those of the author(s) and do not necessarily reflect the views of the National Science Foundation.
\end{acks}

\bibliographystyle{ACM-Reference-Format}
\bibliography{ref/ref,ref/policyHCI,ref/PIprior}


\begin{thebibliography}{98}


\ifx \showCODEN    \undefined \def \showCODEN     #1{\unskip}     \fi
\ifx \showDOI      \undefined \def \showDOI       #1{#1}\fi
\ifx \showISBNx    \undefined \def \showISBNx     #1{\unskip}     \fi
\ifx \showISBNxiii \undefined \def \showISBNxiii  #1{\unskip}     \fi
\ifx \showISSN     \undefined \def \showISSN      #1{\unskip}     \fi
\ifx \showLCCN     \undefined \def \showLCCN      #1{\unskip}     \fi
\ifx \shownote     \undefined \def \shownote      #1{#1}          \fi
\ifx \showarticletitle \undefined \def \showarticletitle #1{#1}   \fi
\ifx \showURL      \undefined \def \showURL       {\relax}        \fi
\providecommand\bibfield[2]{#2}
\providecommand\bibinfo[2]{#2}
\providecommand\natexlab[1]{#1}
\providecommand\showeprint[2][]{arXiv:#2}

\bibitem[Fed(2011)]%
        {FederalRegister_rulemaking}
 \bibinfo{year}{2011}\natexlab{}.
\newblock
\newblock
\urldef\tempurl%
\url{https://www.federalregister.gov/uploads/2011/01/the_rulemaking_process.pdf}
\showURL{%
\tempurl}


\bibitem[US_(2012)]%
        {US_v_Jones}
 \bibinfo{year}{2012}\natexlab{}.
\newblock \bibinfo{title}{United States v. Jones, 565 U.S. 400 (2012)}.
\newblock
\newblock
\urldef\tempurl%
\url{https://supreme.justia.com/cases/federal/us/565/400/}
\showURL{%
\tempurl}


\bibitem[UK_(2017)]%
        {UK_open_policy_toolkit}
 \bibinfo{year}{2017}\natexlab{}.
\newblock
\newblock
\urldef\tempurl%
\url{https://www.gov.uk/guidance/open-policy-making-toolkit}
\showURL{%
\tempurl}


\bibitem[Con(2020)]%
        {Congress_COVID_contact_tracing}
 \bibinfo{year}{2020}\natexlab{}.
\newblock
\newblock
\urldef\tempurl%
\url{https://crsreports.congress.gov/product/pdf/LSB/LSB10511}
\showURL{%
\tempurl}


\bibitem[ftc(2021)]%
        {ftc_workshop_darkpattern}
 \bibinfo{year}{2021}\natexlab{}.
\newblock
\newblock
\urldef\tempurl%
\url{https://www.ftc.gov/system/files/documents/public_events/1586943/dark_patterns-workshop-bios.pdf}
\showURL{%
\tempurl}


\bibitem[sec(2022)]%
        {securityBreach_notification_laws_2022}
 \bibinfo{year}{2022}\natexlab{}.
\newblock
\newblock
\urldef\tempurl%
\url{https://www.ncsl.org/technology-and-communication/security-breach-notification-laws}
\showURL{%
\tempurl}


\bibitem[Accenture(2023)]%
        {AccentureVRgov_2023}
\bibfield{author}{\bibinfo{person}{Accenture}.}
  \bibinfo{year}{2023}\natexlab{}.
\newblock \bibinfo{title}{Government enters the metaverse}.
\newblock
\newblock
\urldef\tempurl%
\url{https://www.accenture.com/us-en/insightsnew/us-federal-government/technology-vision-2022}
\showURL{%
\tempurl}


\bibitem[Ackerman(2000)]%
        {feasbility_gap}
\bibfield{author}{\bibinfo{person}{Mark~S Ackerman}.}
  \bibinfo{year}{2000}\natexlab{}.
\newblock \showarticletitle{The intellectual challenge of CSCW: the gap between
  social requirements and technical feasibility}.
\newblock \bibinfo{journal}{\emph{Human--Computer Interaction}}
  \bibinfo{volume}{15}, \bibinfo{number}{2-3} (\bibinfo{year}{2000}),
  \bibinfo{pages}{179--203}.
\newblock


\bibitem[Albert and Grimmelmann(2023)]%
        {DotheRightThing_CACM2023}
\bibfield{author}{\bibinfo{person}{Kendra Albert} {and} \bibinfo{person}{James
  Grimmelmann}.} \bibinfo{year}{2023}\natexlab{}.
\newblock \showarticletitle{Do the Right Thing}.
\newblock \bibinfo{journal}{\emph{Commun. ACM}} \bibinfo{volume}{66},
  \bibinfo{number}{5} (\bibinfo{year}{2023}), \bibinfo{pages}{18--20}.
\newblock


\bibitem[Anh(2023)]%
        {CHI23WS:Tran_adversarial}
\bibfield{author}{\bibinfo{person}{Tran Anh}.} \bibinfo{year}{2023}\natexlab{}.
\newblock \showarticletitle{Adversarial Engagements Between Technology and
  Policy}. In \bibinfo{booktitle}{\emph{Workshop on Designing Technology and
  Policy Simultaneously: Towards A Research Agenda and New Practice}}.
\newblock


\bibitem[Bamberger and Mulligan(2011)]%
        {bamberger2011privacy}
\bibfield{author}{\bibinfo{person}{Kenneth~A Bamberger} {and}
  \bibinfo{person}{Deirdre~K Mulligan}.} \bibinfo{year}{2011}\natexlab{}.
\newblock \showarticletitle{Privacy on the Books and on the Ground}.
\newblock \bibinfo{journal}{\emph{Stanford Law Review}} (\bibinfo{year}{2011}),
  \bibinfo{pages}{247--315}.
\newblock


\bibitem[Bariso(2016)]%
        {Uber_breaks_Law_INC15}
\bibfield{author}{\bibinfo{person}{Justin Bariso}.}
  \bibinfo{year}{2016}\natexlab{}.
\newblock \bibinfo{title}{Why does uber keep breaking the law? because they're
  disrupting, of course}.
\newblock
\newblock
\urldef\tempurl%
\url{https://www.inc.com/justin-bariso/why-does-uber-keep-breaking-the-law-because-theyre-disrupting-of-course.html}
\showURL{%
\tempurl}


\bibitem[Barocas et~al\mbox{.}(2017)]%
        {solon2017fairness}
\bibfield{author}{\bibinfo{person}{Solon Barocas}, \bibinfo{person}{Moritz
  Hardt}, {and} \bibinfo{person}{Arvind Narayanan}.}
  \bibinfo{year}{2017}\natexlab{}.
\newblock \showarticletitle{Fairness in machine learning}.
\newblock \bibinfo{journal}{\emph{Nips tutorial}}  \bibinfo{volume}{1}
  (\bibinfo{year}{2017}), \bibinfo{pages}{2017}.
\newblock


\bibitem[Blevis(2007)]%
        {blevis2007sustainable}
\bibfield{author}{\bibinfo{person}{Eli Blevis}.}
  \bibinfo{year}{2007}\natexlab{}.
\newblock \showarticletitle{Sustainable interaction design: invention \&
  disposal, renewal \& reuse}. In \bibinfo{booktitle}{\emph{Proceedings of the
  SIGCHI conference on Human factors in computing systems}}.
  \bibinfo{pages}{503--512}.
\newblock


\bibitem[B{\o}dker(2015)]%
        {bodker2015third}
\bibfield{author}{\bibinfo{person}{Susanne B{\o}dker}.}
  \bibinfo{year}{2015}\natexlab{}.
\newblock \showarticletitle{Third-wave HCI, 10 years later---participation and
  sharing}.
\newblock \bibinfo{journal}{\emph{interactions}} \bibinfo{volume}{22},
  \bibinfo{number}{5} (\bibinfo{year}{2015}), \bibinfo{pages}{24--31}.
\newblock


\bibitem[Bommasani et~al\mbox{.}(2021)]%
        {bommasani2021opportunities}
\bibfield{author}{\bibinfo{person}{Rishi Bommasani}, \bibinfo{person}{Drew~A.
  Hudson}, \bibinfo{person}{Ehsan Adeli}, \bibinfo{person}{Russ Altman},
  \bibinfo{person}{Simran Arora}, \bibinfo{person}{Sydney von Arx},
  \bibinfo{person}{Michael~S. Bernstein}, \bibinfo{person}{Jeannette Bohg},
  \bibinfo{person}{Antoine Bosselut}, \bibinfo{person}{Emma Brunskill},
  \bibinfo{person}{Erik Brynjolfsson}, \bibinfo{person}{Shyamal Buch},
  \bibinfo{person}{Dallas Card}, \bibinfo{person}{Rodrigo Castellon},
  \bibinfo{person}{Niladri Chatterji}, \bibinfo{person}{Annie Chen},
  \bibinfo{person}{Kathleen Creel}, \bibinfo{person}{Jared~Quincy Davis},
  \bibinfo{person}{Dora Demszky}, \bibinfo{person}{Chris Donahue},
  \bibinfo{person}{Moussa Doumbouya}, \bibinfo{person}{Esin Durmus},
  \bibinfo{person}{Stefano Ermon}, \bibinfo{person}{John Etchemendy},
  \bibinfo{person}{Kawin Ethayarajh}, \bibinfo{person}{Li Fei-Fei},
  \bibinfo{person}{Chelsea Finn}, \bibinfo{person}{Trevor Gale},
  \bibinfo{person}{Lauren Gillespie}, \bibinfo{person}{Karan Goel},
  \bibinfo{person}{Noah Goodman}, \bibinfo{person}{Shelby Grossman},
  \bibinfo{person}{Neel Guha}, \bibinfo{person}{Tatsunori Hashimoto},
  \bibinfo{person}{Peter Henderson}, \bibinfo{person}{John Hewitt},
  \bibinfo{person}{Daniel~E. Ho}, \bibinfo{person}{Jenny Hong},
  \bibinfo{person}{Kyle Hsu}, \bibinfo{person}{Jing Huang},
  \bibinfo{person}{Thomas Icard}, \bibinfo{person}{Saahil Jain},
  \bibinfo{person}{Dan Jurafsky}, \bibinfo{person}{Pratyusha Kalluri},
  \bibinfo{person}{Siddharth Karamcheti}, \bibinfo{person}{Geoff Keeling},
  \bibinfo{person}{Fereshte Khani}, \bibinfo{person}{Omar Khattab},
  \bibinfo{person}{Pang~Wei Kohd}, \bibinfo{person}{Mark Krass},
  \bibinfo{person}{Ranjay Krishna}, \bibinfo{person}{Rohith Kuditipudi},
  \bibinfo{person}{Ananya Kumar}, \bibinfo{person}{Faisal Ladhak},
  \bibinfo{person}{Mina Lee}, \bibinfo{person}{Tony Lee}, \bibinfo{person}{Jure
  Leskovec}, \bibinfo{person}{Isabelle Levent}, \bibinfo{person}{Xiang~Lisa
  Li}, \bibinfo{person}{Xuechen Li}, \bibinfo{person}{Tengyu Ma},
  \bibinfo{person}{Ali Malik}, \bibinfo{person}{Christopher~D. Manning},
  \bibinfo{person}{Suvir Mirchandani}, \bibinfo{person}{Eric Mitchell},
  \bibinfo{person}{Zanele Munyikwa}, \bibinfo{person}{Suraj Nair},
  \bibinfo{person}{Avanika Narayan}, \bibinfo{person}{Deepak Narayanan},
  \bibinfo{person}{Ben Newman}, \bibinfo{person}{Allen Nie},
  \bibinfo{person}{Juan~Carlos Niebles}, \bibinfo{person}{Hamed Nilforoshan},
  \bibinfo{person}{Julian Nyarko}, \bibinfo{person}{Giray Ogut},
  \bibinfo{person}{Laurel Orr}, \bibinfo{person}{Isabel Papadimitriou},
  \bibinfo{person}{Joon~Sung Park}, \bibinfo{person}{Chris Piech},
  \bibinfo{person}{Eva Portelance}, \bibinfo{person}{Christopher Potts},
  \bibinfo{person}{Aditi Raghunathan}, \bibinfo{person}{Rob Reich},
  \bibinfo{person}{Hongyu Ren}, \bibinfo{person}{Frieda Rong},
  \bibinfo{person}{Yusuf Roohani}, \bibinfo{person}{Camilo Ruiz},
  \bibinfo{person}{Jack Ryan}, \bibinfo{person}{Christopher Ré},
  \bibinfo{person}{Dorsa Sadigh}, \bibinfo{person}{Shiori Sagawa},
  \bibinfo{person}{Keshav Santhanam}, \bibinfo{person}{Andy Shih},
  \bibinfo{person}{Krishnan Srinivasan}, \bibinfo{person}{Alex Tamkin},
  \bibinfo{person}{Rohan Taori}, \bibinfo{person}{Armin~W. Thomas},
  \bibinfo{person}{Florian Tramèr}, \bibinfo{person}{Rose~E. Wang},
  \bibinfo{person}{William Wang}, \bibinfo{person}{Bohan Wu},
  \bibinfo{person}{Jiajun Wu}, \bibinfo{person}{Yuhuai Wu},
  \bibinfo{person}{Sang~Michael Xie}, \bibinfo{person}{Michihiro Yasunaga},
  \bibinfo{person}{Jiaxuan You}, \bibinfo{person}{Matei Zaharia},
  \bibinfo{person}{Michael Zhang}, \bibinfo{person}{Tianyi Zhang},
  \bibinfo{person}{Xikun Zhang}, \bibinfo{person}{Yuhui Zhang},
  \bibinfo{person}{Lucia Zheng}, \bibinfo{person}{Kaitlyn Zhou}, {and}
  \bibinfo{person}{Percy Liang}.} \bibinfo{year}{2021}\natexlab{}.
\newblock \bibinfo{title}{On the Opportunities and Risks of Foundation Models}.
\newblock
\newblock
\showeprint[arxiv]{2108.07258}~[cs.LG]


\bibitem[Bork(2019)]%
        {bork2019evolutionary}
\bibfield{author}{\bibinfo{person}{Karrigan~S Bork}.}
  \bibinfo{year}{2019}\natexlab{}.
\newblock \showarticletitle{An evolutionary theory of administrative law}.
\newblock \bibinfo{journal}{\emph{SMU L. Rev.}}  \bibinfo{volume}{72}
  (\bibinfo{year}{2019}), \bibinfo{pages}{81}.
\newblock


\bibitem[Brubaker and Callison-Burch(2016)]%
        {FacebookPolicy_CHI16}
\bibfield{author}{\bibinfo{person}{Jed~R. Brubaker} {and}
  \bibinfo{person}{Vanessa Callison-Burch}.} \bibinfo{year}{2016}\natexlab{}.
\newblock \showarticletitle{Legacy Contact: Designing and Implementing
  Post-Mortem Stewardship at Facebook}. In
  \bibinfo{booktitle}{\emph{Proceedings of the 2016 CHI Conference on Human
  Factors in Computing Systems}} (San Jose, California, USA)
  \emph{(\bibinfo{series}{CHI '16})}. \bibinfo{publisher}{Association for
  Computing Machinery}, \bibinfo{address}{New York, NY, USA},
  \bibinfo{pages}{2908–2919}.
\newblock
\showISBNx{9781450333627}
\urldef\tempurl%
\url{https://doi.org/10.1145/2858036.2858254}
\showDOI{\tempurl}


\bibitem[Burgess(2023)]%
        {CriminalGPT_Wired2023}
\bibfield{author}{\bibinfo{person}{Matt Burgess}.}
  \bibinfo{year}{2023}\natexlab{}.
\newblock \bibinfo{title}{Criminals Have Created Their Own ChatGPT Clones}.
\newblock
\newblock
\urldef\tempurl%
\url{https://www.wired.com/story/chatgpt-scams-fraudgpt-wormgpt-crime/}
\showURL{%
\tempurl}


\bibitem[Centivany(2016)]%
        {policy-generativity-Centivany-CSCW16}
\bibfield{author}{\bibinfo{person}{Alissa Centivany}.}
  \bibinfo{year}{2016}\natexlab{}.
\newblock \showarticletitle{Policy as Embedded Generativity: A Case Study of
  the Emergence and Evolution of HathiTrust}. In
  \bibinfo{booktitle}{\emph{Proceedings of the 19th ACM Conference on
  Computer-Supported Cooperative Work \& Social Computing}} (San Francisco,
  California, USA) \emph{(\bibinfo{series}{CSCW '16})}.
  \bibinfo{publisher}{Association for Computing Machinery},
  \bibinfo{address}{New York, NY, USA}, \bibinfo{pages}{926–940}.
\newblock
\showISBNx{9781450335928}
\urldef\tempurl%
\url{https://doi.org/10.1145/2818048.2820069}
\showDOI{\tempurl}


\bibitem[Chasalow and Levy(2021)]%
        {chasalow2021representativeness}
\bibfield{author}{\bibinfo{person}{Kyla Chasalow} {and} \bibinfo{person}{Karen
  Levy}.} \bibinfo{year}{2021}\natexlab{}.
\newblock \showarticletitle{Representativeness in statistics, politics, and
  machine learning}. In \bibinfo{booktitle}{\emph{Proceedings of the 2021 ACM
  Conference on Fairness, Accountability, and Transparency}}.
  \bibinfo{pages}{77--89}.
\newblock


\bibitem[Dal~B{\'o}(2006)]%
        {regulatory_capture2006}
\bibfield{author}{\bibinfo{person}{Ernesto Dal~B{\'o}}.}
  \bibinfo{year}{2006}\natexlab{}.
\newblock \showarticletitle{Regulatory capture: A review}.
\newblock \bibinfo{journal}{\emph{Oxford review of economic policy}}
  \bibinfo{volume}{22}, \bibinfo{number}{2} (\bibinfo{year}{2006}),
  \bibinfo{pages}{203--225}.
\newblock


\bibitem[Dardaman and Gupta(2023)]%
        {CHI23WS:BCG_Forecasting}
\bibfield{author}{\bibinfo{person}{Emily Dardaman} {and}
  \bibinfo{person}{Abhishek Gupta}.} \bibinfo{year}{2023}\natexlab{}.
\newblock \showarticletitle{Asking Better Questions -- The Art and Science of
  Forecasting: A mechanism for truer answers to high-stakes questions}. In
  \bibinfo{booktitle}{\emph{Workshop on Designing Technology and Policy
  Simultaneously: Towards A Research Agenda and New Practice}}.
\newblock
\showeprint[arxiv]{2303.18006}~[cs.CY]


\bibitem[Davis et~al\mbox{.}(2012)]%
        {OccupyCHI_USpolicy_CHI12panel}
\bibfield{author}{\bibinfo{person}{Janet Davis}, \bibinfo{person}{Harry
  Hochheiser}, \bibinfo{person}{Juan~Pablo Hourcade}, \bibinfo{person}{Jeff
  Johnson}, \bibinfo{person}{Lisa Nathan}, {and} \bibinfo{person}{Janice
  Tsai}.} \bibinfo{year}{2012}\natexlab{}.
\newblock \showarticletitle{Occupy CHI! Engaging U.S. Policymakers}. In
  \bibinfo{booktitle}{\emph{CHI '12 Extended Abstracts on Human Factors in
  Computing Systems}} (Austin, Texas, USA) \emph{(\bibinfo{series}{CHI EA
  '12})}. \bibinfo{publisher}{Association for Computing Machinery},
  \bibinfo{address}{New York, NY, USA}, \bibinfo{pages}{1139–1142}.
\newblock
\showISBNx{9781450310161}
\urldef\tempurl%
\url{https://doi.org/10.1145/2212776.2212406}
\showDOI{\tempurl}
\newblock
\shownote{Note: Panel slides can be found at
  \url{https://web.archive.org/web/20170829055308/http://www.cs.grinnell.edu/~davisjan/chi-us-public-policy/chi2012-panel.pdf}.
  Accessed: 2023-07-31}.


\bibitem[Delgado et~al\mbox{.}(2023)]%
        {delgado2023participatory}
\bibfield{author}{\bibinfo{person}{Fernando Delgado}, \bibinfo{person}{Stephen
  Yang}, \bibinfo{person}{Michael Madaio}, {and} \bibinfo{person}{Qian Yang}.}
  \bibinfo{year}{2023}\natexlab{}.
\newblock \showarticletitle{The Participatory Turn in AI Design: Theoretical
  Foundations and the Current State of Practice}. In
  \bibinfo{booktitle}{\emph{Proceedings of the 3rd ACM Conference on Equity and
  Access in Algorithms, Mechanisms, and Optimization}}. \bibinfo{pages}{1--23}.
\newblock


\bibitem[DiSalvo et~al\mbox{.}(2014)]%
        {disalvo2014making}
\bibfield{author}{\bibinfo{person}{Carl DiSalvo}, \bibinfo{person}{Jonathan
  Lukens}, \bibinfo{person}{Thomas Lodato}, \bibinfo{person}{Tom Jenkins},
  {and} \bibinfo{person}{Tanyoung Kim}.} \bibinfo{year}{2014}\natexlab{}.
\newblock \showarticletitle{Making public things: how HCI design can express
  matters of concern}. In \bibinfo{booktitle}{\emph{Proceedings of the SIGCHI
  Conference on Human factors in Computing Systems}}.
  \bibinfo{pages}{2397--2406}.
\newblock


\bibitem[Dombrowski et~al\mbox{.}(2016)]%
        {Dombrowski_SocialJusticeIXD_DIS16}
\bibfield{author}{\bibinfo{person}{Lynn Dombrowski}, \bibinfo{person}{Ellie
  Harmon}, {and} \bibinfo{person}{Sarah Fox}.} \bibinfo{year}{2016}\natexlab{}.
\newblock \showarticletitle{Social Justice-Oriented Interaction Design:
  Outlining Key Design Strategies and Commitments}. In
  \bibinfo{booktitle}{\emph{Proceedings of the 2016 ACM Conference on Designing
  Interactive Systems}} (Brisbane, QLD, Australia) \emph{(\bibinfo{series}{DIS
  '16})}. \bibinfo{publisher}{Association for Computing Machinery},
  \bibinfo{address}{New York, NY, USA}, \bibinfo{pages}{656–671}.
\newblock
\showISBNx{9781450340311}
\urldef\tempurl%
\url{https://doi.org/10.1145/2901790.2901861}
\showDOI{\tempurl}


\bibitem[Dorst and Cross(2001)]%
        {dorst2001creativity}
\bibfield{author}{\bibinfo{person}{Kees Dorst} {and} \bibinfo{person}{Nigel
  Cross}.} \bibinfo{year}{2001}\natexlab{}.
\newblock \showarticletitle{Creativity in the design process: co-evolution of
  problem--solution}.
\newblock \bibinfo{journal}{\emph{Design studies}} \bibinfo{volume}{22},
  \bibinfo{number}{5} (\bibinfo{year}{2001}), \bibinfo{pages}{425--437}.
\newblock


\bibitem[Dym et~al\mbox{.}(2022)]%
        {pillowfort-platform-design-and-policy-chi22}
\bibfield{author}{\bibinfo{person}{Brianna Dym}, \bibinfo{person}{Namita
  Pasupuleti}, {and} \bibinfo{person}{Casey Fiesler}.}
  \bibinfo{year}{2022}\natexlab{}.
\newblock \showarticletitle{Building a Pillowfort: Political Tensions in
  Platform Design and Policy}.
\newblock \bibinfo{journal}{\emph{Proc. ACM Hum.-Comput. Interact.}}
  \bibinfo{volume}{6}, \bibinfo{number}{GROUP}, Article \bibinfo{articleno}{16}
  (\bibinfo{date}{jan} \bibinfo{year}{2022}), \bibinfo{numpages}{23}~pages.
\newblock
\urldef\tempurl%
\url{https://doi.org/10.1145/3492835}
\showDOI{\tempurl}


\bibitem[Edwards et~al\mbox{.}(2007)]%
        {Jackson2007understanding_infra}
\bibfield{author}{\bibinfo{person}{Paul~N Edwards}, \bibinfo{person}{Steven~J
  Jackson}, \bibinfo{person}{Geoffrey~C Bowker}, {and}
  \bibinfo{person}{Cory~Philip Knobel}.} \bibinfo{year}{2007}\natexlab{}.
\newblock \showarticletitle{Understanding infrastructure: Dynamics, tensions,
  and design}.
\newblock  (\bibinfo{year}{2007}).
\newblock


\bibitem[Edwards et~al\mbox{.}(2010)]%
        {InfraProblemInHCI_CHI10}
\bibfield{author}{\bibinfo{person}{W.~Keith Edwards}, \bibinfo{person}{Mark~W.
  Newman}, {and} \bibinfo{person}{Erika~Shehan Poole}.}
  \bibinfo{year}{2010}\natexlab{}.
\newblock \showarticletitle{The Infrastructure Problem in HCI}. In
  \bibinfo{booktitle}{\emph{Proceedings of the SIGCHI Conference on Human
  Factors in Computing Systems}} (Atlanta, Georgia, USA)
  \emph{(\bibinfo{series}{CHI '10})}. \bibinfo{publisher}{Association for
  Computing Machinery}, \bibinfo{address}{New York, NY, USA},
  \bibinfo{pages}{423–432}.
\newblock
\showISBNx{9781605589299}
\urldef\tempurl%
\url{https://doi.org/10.1145/1753326.1753390}
\showDOI{\tempurl}


\bibitem[Forlizzi(2018a)]%
        {UCDdead}
\bibfield{author}{\bibinfo{person}{Jodi Forlizzi}.}
  \bibinfo{year}{2018}\natexlab{a}.
\newblock \showarticletitle{Moving beyond user-centered design}.
\newblock \bibinfo{journal}{\emph{interactions}} \bibinfo{volume}{25},
  \bibinfo{number}{5} (\bibinfo{year}{2018}), \bibinfo{pages}{22--23}.
\newblock


\bibitem[Forlizzi(2018b)]%
        {forlizzi2018moving}
\bibfield{author}{\bibinfo{person}{Jodi Forlizzi}.}
  \bibinfo{year}{2018}\natexlab{b}.
\newblock \showarticletitle{Moving beyond user-centered design}.
\newblock \bibinfo{journal}{\emph{interactions}} \bibinfo{volume}{25},
  \bibinfo{number}{5} (\bibinfo{year}{2018}), \bibinfo{pages}{22--23}.
\newblock


\bibitem[Gellman(2022)]%
        {gellman2022fair}
\bibfield{author}{\bibinfo{person}{Robert Gellman}.}
  \bibinfo{year}{2022}\natexlab{}.
\newblock \showarticletitle{Fair Information Practices: A Basic History-Version
  2.22}.
\newblock \bibinfo{journal}{\emph{Available at SSRN}} (\bibinfo{year}{2022}).
\newblock


\bibitem[Gilbert et~al\mbox{.}(2022)]%
        {Gilbert-sociotechnical-specification}
\bibfield{author}{\bibinfo{person}{Thomas~Krendl Gilbert},
  \bibinfo{person}{Aaron~J Snoswell}, \bibinfo{person}{Michael Dennis},
  \bibinfo{person}{Rowan Mcallister}, {and} \bibinfo{person}{Cathy Wu}.}
  \bibinfo{year}{2022}\natexlab{}.
\newblock \showarticletitle{Sociotechnical Specification for the Broader
  Impacts of Autonomous Vehicles}.
\newblock \bibinfo{journal}{\emph{ICRA’22 workshop Fresh Perspectives on the
  Future of Autonomous Driving}}.
\newblock


\bibitem[Gray(2022)]%
        {gray2022arizona_court}
\bibfield{author}{\bibinfo{person}{Colin~M Gray}.}
  \bibinfo{year}{2022}\natexlab{}.
\newblock \showarticletitle{Expert Report of Colin M. Gray, Ph. D. on Dark
  Patterns (Public Redacted Version)}.
\newblock \bibinfo{journal}{\emph{STATE OF ARIZONA, ex rel. MARK BRNOVICH,
  Attorney General, Plaintiff, v. Google LLC}} (\bibinfo{year}{2022}).
\newblock
\urldef\tempurl%
\url{https://www.azag.gov/sites/default/files/2022-09/Expert%20Report%20of%20Colin%20M.%20Gray%2C%20Ph.D..pdf}
\showURL{%
\tempurl}


\bibitem[Gray et~al\mbox{.}(2018)]%
        {gray_darkpattern_CHI18}
\bibfield{author}{\bibinfo{person}{Colin~M Gray}, \bibinfo{person}{Yubo Kou},
  \bibinfo{person}{Bryan Battles}, \bibinfo{person}{Joseph Hoggatt}, {and}
  \bibinfo{person}{Austin~L Toombs}.} \bibinfo{year}{2018}\natexlab{}.
\newblock \showarticletitle{The dark (patterns) side of UX design}. In
  \bibinfo{booktitle}{\emph{Proceedings of the 2018 CHI conference on human
  factors in computing systems}}. \bibinfo{pages}{1--14}.
\newblock


\bibitem[Haase and Laursen(2019)]%
        {haase2019meaning}
\bibfield{author}{\bibinfo{person}{Louise~M{\o}ller Haase} {and}
  \bibinfo{person}{Linda~Nhu Laursen}.} \bibinfo{year}{2019}\natexlab{}.
\newblock \showarticletitle{Meaning frames: The structure of problem frames and
  solution frames}.
\newblock \bibinfo{journal}{\emph{Design Issues}} \bibinfo{volume}{35},
  \bibinfo{number}{3} (\bibinfo{year}{2019}), \bibinfo{pages}{20--34}.
\newblock


\bibitem[Haimson et~al\mbox{.}(2021)]%
        {haimson2021disproportionate}
\bibfield{author}{\bibinfo{person}{Oliver~L Haimson}, \bibinfo{person}{Daniel
  Delmonaco}, \bibinfo{person}{Peipei Nie}, {and} \bibinfo{person}{Andrea
  Wegner}.} \bibinfo{year}{2021}\natexlab{}.
\newblock \showarticletitle{Disproportionate removals and differing content
  moderation experiences for conservative, transgender, and black social media
  users: Marginalization and moderation gray areas}.
\newblock \bibinfo{journal}{\emph{Proceedings of the ACM on Human-Computer
  Interaction}} \bibinfo{volume}{5}, \bibinfo{number}{CSCW2}
  (\bibinfo{year}{2021}), \bibinfo{pages}{1--35}.
\newblock


\bibitem[Hayes(2002)]%
        {hayes2002policy_change_limits_incrementalism}
\bibfield{author}{\bibinfo{person}{Michael~T Hayes}.}
  \bibinfo{year}{2002}\natexlab{}.
\newblock \bibinfo{booktitle}{\emph{The limits of policy change:
  Incrementalism, worldview, and the rule of law}}.
\newblock \bibinfo{publisher}{Georgetown University Press}.
\newblock


\bibitem[Hickton(2023)]%
        {chatGPT_quickRegulation}
\bibfield{author}{\bibinfo{person}{David Hickton}.}
  \bibinfo{year}{2023}\natexlab{}.
\newblock \bibinfo{title}{Chatgpt could transform society and its risks require
  quick regulation}.
\newblock
\newblock
\urldef\tempurl%
\url{https://thehill.com/opinion/technology/3812597-chatgpt-could-transform-society-and-its-risks-require-quick-regulation/}
\showURL{%
\tempurl}


\bibitem[Hong(2023)]%
        {HongFATEPrivacyPolicy_CACM2023}
\bibfield{author}{\bibinfo{person}{Jason Hong}.}
  \bibinfo{year}{2023}\natexlab{}.
\newblock \bibinfo{title}{What Can the FATE Community Learn from the Successes
  and Failures in Privacy?}
\newblock
  \bibinfo{howpublished}{\url{https://cacm.acm.org/blogs/blog-cacm/271991-what-can-the-fate-community-learn-from-the-successes-and-failures-in-privacy/fulltext}}.
\newblock
\newblock
\shownote{Accessed: 2023-07-29}.


\bibitem[Jackson et~al\mbox{.}(2014)]%
        {policyknot-CSCW14}
\bibfield{author}{\bibinfo{person}{Steven~J. Jackson},
  \bibinfo{person}{Tarleton Gillespie}, {and} \bibinfo{person}{Sandy Payette}.}
  \bibinfo{year}{2014}\natexlab{}.
\newblock \showarticletitle{The Policy Knot: Re-Integrating Policy, Practice
  and Design in CSCW Studies of Social Computing}. In
  \bibinfo{booktitle}{\emph{Proceedings of the 17th ACM Conference on Computer
  Supported Cooperative Work \& Social Computing}} (Baltimore, Maryland, USA)
  \emph{(\bibinfo{series}{CSCW '14})}. \bibinfo{publisher}{Association for
  Computing Machinery}, \bibinfo{address}{New York, NY, USA},
  \bibinfo{pages}{588–602}.
\newblock
\showISBNx{9781450325400}
\urldef\tempurl%
\url{https://doi.org/10.1145/2531602.2531674}
\showDOI{\tempurl}


\bibitem[Jantsch(1972)]%
        {jantsch1972inter_transdisciplinary}
\bibfield{author}{\bibinfo{person}{Erich Jantsch}.}
  \bibinfo{year}{1972}\natexlab{}.
\newblock \showarticletitle{Inter-and transdisciplinary university: A systems
  approach to education and innovation}.
\newblock \bibinfo{journal}{\emph{Higher education}} \bibinfo{volume}{1},
  \bibinfo{number}{1} (\bibinfo{year}{1972}), \bibinfo{pages}{7--37}.
\newblock


\bibitem[Jasanoff(2014)]%
        {jasanoff2014serviceable}
\bibfield{author}{\bibinfo{person}{Sheila Jasanoff}.}
  \bibinfo{year}{2014}\natexlab{}.
\newblock \showarticletitle{Serviceable truths: Science for action in law and
  policy}.
\newblock \bibinfo{journal}{\emph{Tex. L. Rev.}}  \bibinfo{volume}{93}
  (\bibinfo{year}{2014}), \bibinfo{pages}{1723}.
\newblock


\bibitem[Jin(2023)]%
        {CHI23WS:Jin_FAccT_Criminal}
\bibfield{author}{\bibinfo{person}{Angela Jin}.}
  \bibinfo{year}{2023}\natexlab{}.
\newblock \showarticletitle{{Opportunities for Coordinating Policy and
  Technology Design for Algorithmic Accountability: Insights from the U.S.
  Criminal Legal System}}. In \bibinfo{booktitle}{\emph{Workshop on Designing
  Technology and Policy Simultaneously: Towards A Research Agenda and New
  Practice}}.
\newblock


\bibitem[Junginger(2013)]%
        {Junginger2013-matters-of-design-in-policy}
\bibfield{author}{\bibinfo{person}{Sabine Junginger}.}
  \bibinfo{year}{2013}\natexlab{}.
\newblock \showarticletitle{Design and Innovation in the Public Sector: Matters
  of Design in Policy-Making and Policy Implementation}.
\newblock \bibinfo{journal}{\emph{Annual Review of Policy Design}}
  \bibinfo{volume}{1} (\bibinfo{year}{2013}).
\newblock
Issue 1.
\urldef\tempurl%
\url{https://ojs.unbc.ca/index.php/design/article/view/542}
\showURL{%
\tempurl}


\bibitem[Junginger(2016a)]%
        {junginger-book}
\bibfield{author}{\bibinfo{person}{Sabine Junginger}.}
  \bibinfo{year}{2016}\natexlab{a}.
\newblock \bibinfo{booktitle}{\emph{Transforming Public Services by Design:
  Re-orienting policies, organizations and services around people}}.
\newblock \bibinfo{publisher}{Taylor \& Francis}.
\newblock


\bibitem[Junginger(2016b)]%
        {junginger-book-USPS}
\bibfield{author}{\bibinfo{person}{Sabine Junginger}.}
  \bibinfo{year}{2016}\natexlab{b}.
\newblock \showarticletitle{The USPS domestic mail manual transformation
  project}.
\newblock In \bibinfo{booktitle}{\emph{Transforming Public Services by Design:
  Re-orienting policies, organizations and services around people}}.
  \bibinfo{publisher}{Taylor \& Francis}.
\newblock


\bibitem[Katz(2021)]%
        {consequence-scanning-Salesforce}
\bibfield{author}{\bibinfo{person}{Rob Katz}.} \bibinfo{year}{2021}\natexlab{}.
\newblock \bibinfo{title}{How to run a consequence Scanning workshop}.
\newblock
\newblock
\urldef\tempurl%
\url{https://medium.com/salesforce-ux/how-to-run-a-consequence-scanning-workshop-4b14792ea987}
\showURL{%
\tempurl}


\bibitem[Kessler(2013)]%
        {Uber_breaks_Law_FastCompany13}
\bibfield{author}{\bibinfo{person}{Sarah Kessler}.}
  \bibinfo{year}{2013}\natexlab{}.
\newblock \bibinfo{title}{Uber: When innovation outpaces the law}.
\newblock
\newblock
\urldef\tempurl%
\url{https://www.fastcompany.com/3001169/uber-when-innovation-outpaces-law}
\showURL{%
\tempurl}


\bibitem[Kingdon and Stano(1984)]%
        {kingdon1984windows_of_opportunity}
\bibfield{author}{\bibinfo{person}{John~W Kingdon} {and} \bibinfo{person}{Eric
  Stano}.} \bibinfo{year}{1984}\natexlab{}.
\newblock \bibinfo{booktitle}{\emph{Agendas, alternatives, and public
  policies}}. Vol.~\bibinfo{volume}{45}.
\newblock \bibinfo{publisher}{Little, Brown Boston}.
\newblock


\bibitem[Kirkham(2023)]%
        {CHI23WS:ResearchThroughLitigation}
\bibfield{author}{\bibinfo{person}{Reuben Kirkham}.}
  \bibinfo{year}{2023}\natexlab{}.
\newblock \showarticletitle{{(Legal Design) Research through Litigation}}. In
  \bibinfo{booktitle}{\emph{Workshop on Designing Technology and Policy
  Simultaneously: Towards A Research Agenda and New Practice}}.
\newblock
\showeprint[arxiv]{2303.14336}~[cs.HC]


\bibitem[Klein(1990)]%
        {klein1990interdisciplinarity}
\bibfield{author}{\bibinfo{person}{Julie~Thompson Klein}.}
  \bibinfo{year}{1990}\natexlab{}.
\newblock \bibinfo{booktitle}{\emph{Interdisciplinarity: History, theory, and
  practice}}.
\newblock \bibinfo{publisher}{Wayne state university press}.
\newblock


\bibitem[Kriebel et~al\mbox{.}(2001)]%
        {kriebel2001precautionary}
\bibfield{author}{\bibinfo{person}{David Kriebel}, \bibinfo{person}{Joel
  Tickner}, \bibinfo{person}{Paul Epstein}, \bibinfo{person}{John Lemons},
  \bibinfo{person}{Richard Levins}, \bibinfo{person}{Edward~L Loechler},
  \bibinfo{person}{Margaret Quinn}, \bibinfo{person}{Ruthann Rudel},
  \bibinfo{person}{Ted Schettler}, {and} \bibinfo{person}{Michael Stoto}.}
  \bibinfo{year}{2001}\natexlab{}.
\newblock \showarticletitle{The precautionary principle in environmental
  science.}
\newblock \bibinfo{journal}{\emph{Environmental health perspectives}}
  \bibinfo{volume}{109}, \bibinfo{number}{9} (\bibinfo{year}{2001}),
  \bibinfo{pages}{871--876}.
\newblock


\bibitem[Laufer et~al\mbox{.}(2022)]%
        {laufer2022four}
\bibfield{author}{\bibinfo{person}{Benjamin Laufer}, \bibinfo{person}{Sameer
  Jain}, \bibinfo{person}{A~Feder Cooper}, \bibinfo{person}{Jon Kleinberg},
  {and} \bibinfo{person}{Hoda Heidari}.} \bibinfo{year}{2022}\natexlab{}.
\newblock \showarticletitle{Four years of FAccT: A reflexive, mixed-methods
  analysis of research contributions, shortcomings, and future prospects}. In
  \bibinfo{booktitle}{\emph{Proceedings of the 2022 ACM Conference on Fairness,
  Accountability, and Transparency}}. \bibinfo{pages}{401--426}.
\newblock


\bibitem[Lazar(2010)]%
        {Lazar-interacting-with-public-policy}
\bibfield{author}{\bibinfo{person}{Jonathan Lazar}.}
  \bibinfo{year}{2010}\natexlab{}.
\newblock \showarticletitle{Interacting with Public Policy}.
\newblock \bibinfo{journal}{\emph{Interactions}} \bibinfo{volume}{17},
  \bibinfo{number}{1} (\bibinfo{date}{jan} \bibinfo{year}{2010}),
  \bibinfo{pages}{40–43}.
\newblock
\showISSN{1072-5520}
\urldef\tempurl%
\url{https://doi.org/10.1145/1649475.1649485}
\showDOI{\tempurl}


\bibitem[Lazar(2015)]%
        {Lazar-HCI-policy-Interacitons15-forum}
\bibfield{author}{\bibinfo{person}{Jonathan Lazar}.}
  \bibinfo{year}{2015}\natexlab{}.
\newblock \showarticletitle{Public Policy and HCI: Making an Impact in the
  Future}.
\newblock \bibinfo{journal}{\emph{Interactions}} \bibinfo{volume}{22},
  \bibinfo{number}{5} (\bibinfo{date}{aug} \bibinfo{year}{2015}),
  \bibinfo{pages}{69–71}.
\newblock
\showISSN{1072-5520}
\urldef\tempurl%
\url{https://doi.org/10.1145/2807916}
\showDOI{\tempurl}


\bibitem[Lazar(2017)]%
        {lazar2017_gap_year}
\bibfield{author}{\bibinfo{person}{Jonathan Lazar}.}
  \bibinfo{year}{2017}\natexlab{}.
\newblock \showarticletitle{Let's strengthen the HCI community by taking a gap
  year!}
\newblock \bibinfo{journal}{\emph{Interactions}} \bibinfo{volume}{25},
  \bibinfo{number}{1} (\bibinfo{year}{2017}), \bibinfo{pages}{20--21}.
\newblock


\bibitem[Lazar et~al\mbox{.}(2012)]%
        {Lazer_10year_HCIpulicpolicy2012}
\bibfield{author}{\bibinfo{person}{Jonathan Lazar}, \bibinfo{person}{Julio
  Abascal}, \bibinfo{person}{Janet Davis}, \bibinfo{person}{Vanessa Evers},
  \bibinfo{person}{Jan Gulliksen}, \bibinfo{person}{Joaquim Jorge},
  \bibinfo{person}{Tom McEwan}, \bibinfo{person}{Fabio Patern\`{o}},
  \bibinfo{person}{Hans Persson}, \bibinfo{person}{Raquel Prates},
  \bibinfo{person}{Hans von Axelson}, \bibinfo{person}{Marco Winckler}, {and}
  \bibinfo{person}{Volker Wulf}.} \bibinfo{year}{2012}\natexlab{}.
\newblock \showarticletitle{HCI Public Policy Activities in 2012: A 10-Country
  Discussion}.
\newblock \bibinfo{journal}{\emph{Interactions}} \bibinfo{volume}{19},
  \bibinfo{number}{3} (\bibinfo{date}{may} \bibinfo{year}{2012}),
  \bibinfo{pages}{78–81}.
\newblock
\showISSN{1072-5520}
\urldef\tempurl%
\url{https://doi.org/10.1145/2168931.2168947}
\showDOI{\tempurl}


\bibitem[Lee et~al\mbox{.}(2015)]%
        {lee2015working}
\bibfield{author}{\bibinfo{person}{Min~Kyung Lee}, \bibinfo{person}{Daniel
  Kusbit}, \bibinfo{person}{Evan Metsky}, {and} \bibinfo{person}{Laura
  Dabbish}.} \bibinfo{year}{2015}\natexlab{}.
\newblock \showarticletitle{Working with machines: The impact of algorithmic
  and data-driven management on human workers}. In
  \bibinfo{booktitle}{\emph{Proceedings of the 33rd annual ACM conference on
  human factors in computing systems}}. \bibinfo{pages}{1603--1612}.
\newblock


\bibitem[Lindley et~al\mbox{.}(2017)]%
        {lindley2017anticipating}
\bibfield{author}{\bibinfo{person}{Joseph~Galen Lindley}, \bibinfo{person}{Paul
  Coulton}, \bibinfo{person}{Haider Akmal}, {and}
  \bibinfo{person}{Brandin~Hanson Knowles}.} \bibinfo{year}{2017}\natexlab{}.
\newblock \showarticletitle{Anticipating GDPR in Smart Homes Through Fictional
  Conversational Objects}.
\newblock  (\bibinfo{year}{2017}).
\newblock


\bibitem[Madaio et~al\mbox{.}(2020)]%
        {Fairness_Checklist}
\bibfield{author}{\bibinfo{person}{Michael~A. Madaio}, \bibinfo{person}{Luke
  Stark}, \bibinfo{person}{Jennifer Wortman~Vaughan}, {and}
  \bibinfo{person}{Hanna Wallach}.} \bibinfo{year}{2020}\natexlab{}.
\newblock \showarticletitle{Co-Designing Checklists to Understand
  Organizational Challenges and Opportunities around Fairness in AI}. In
  \bibinfo{booktitle}{\emph{Proceedings of the 2020 CHI Conference on Human
  Factors in Computing Systems}} (Honolulu, HI, USA)
  \emph{(\bibinfo{series}{CHI '20})}. \bibinfo{publisher}{Association for
  Computing Machinery}, \bibinfo{address}{New York, NY, USA},
  \bibinfo{pages}{1–14}.
\newblock
\showISBNx{9781450367080}
\urldef\tempurl%
\url{https://doi.org/10.1145/3313831.3376445}
\showDOI{\tempurl}


\bibitem[Messeri and Vertesi(2015)]%
        {messeri2015greatest}
\bibfield{author}{\bibinfo{person}{Lisa Messeri} {and} \bibinfo{person}{Janet
  Vertesi}.} \bibinfo{year}{2015}\natexlab{}.
\newblock \showarticletitle{The Greatest Missions Never Flown: Anticipatory
  Discourse and the" Projectory" in Technological Communities}.
\newblock \bibinfo{journal}{\emph{Technology and Culture}}
  (\bibinfo{year}{2015}), \bibinfo{pages}{54--85}.
\newblock


\bibitem[Moor(1985)]%
        {moor1985computer}
\bibfield{author}{\bibinfo{person}{James~H Moor}.}
  \bibinfo{year}{1985}\natexlab{}.
\newblock \showarticletitle{What is computer ethics?}
\newblock \bibinfo{journal}{\emph{Metaphilosophy}} \bibinfo{volume}{16},
  \bibinfo{number}{4} (\bibinfo{year}{1985}), \bibinfo{pages}{266--275}.
\newblock


\bibitem[Nathan and Friedman(2010)]%
        {Nathan_Interactions2010_RwandaPolicy}
\bibfield{author}{\bibinfo{person}{Lisa~P. Nathan} {and} \bibinfo{person}{Batya
  Friedman}.} \bibinfo{year}{2010}\natexlab{}.
\newblock \showarticletitle{Interacting with Policy in a Political World:
  Reflections from the Voices from the Rwanda Tribunal Project}.
\newblock \bibinfo{journal}{\emph{Interactions}} \bibinfo{volume}{17},
  \bibinfo{number}{5} (\bibinfo{date}{sep} \bibinfo{year}{2010}),
  \bibinfo{pages}{56–59}.
\newblock
\showISSN{1072-5520}
\urldef\tempurl%
\url{https://doi.org/10.1145/1836216.1836231}
\showDOI{\tempurl}


\bibitem[Newman(2019)]%
        {newman2019found}
\bibfield{author}{\bibinfo{person}{Andy Newman}.}
  \bibinfo{year}{2019}\natexlab{}.
\newblock \showarticletitle{I found work on an Amazon website. I made 97 cents
  an hour}.
\newblock \bibinfo{journal}{\emph{The New York Times}}  \bibinfo{volume}{15}
  (\bibinfo{year}{2019}).
\newblock


\bibitem[Nudzor(2009)]%
        {nudzor2009policy}
\bibfield{author}{\bibinfo{person}{Hope Nudzor}.}
  \bibinfo{year}{2009}\natexlab{}.
\newblock \showarticletitle{What is “policy”, a problem: solving definition
  or a process conceptualisation}.
\newblock \bibinfo{journal}{\emph{Educational futures}} \bibinfo{volume}{2},
  \bibinfo{number}{1} (\bibinfo{year}{2009}), \bibinfo{pages}{85--96}.
\newblock


\bibitem[Perrigo(2023)]%
        {OpenAI_Lobbying_EU_AI}
\bibfield{author}{\bibinfo{person}{Billy Perrigo}.}
  \bibinfo{year}{2023}\natexlab{}.
\newblock \bibinfo{title}{Exclusive: Openai lobbied E.U. to water down AI
  regulation}.
\newblock
\newblock
\urldef\tempurl%
\url{https://time.com/6288245/openai-eu-lobbying-ai-act/}
\showURL{%
\tempurl}


\bibitem[Peters(2021)]%
        {peters2021advanced}
\bibfield{author}{\bibinfo{person}{B~Guy Peters}.}
  \bibinfo{year}{2021}\natexlab{}.
\newblock \bibinfo{booktitle}{\emph{Advanced introduction to public policy}}.
\newblock \bibinfo{publisher}{Edward Elgar Publishing}.
\newblock


\bibitem[Ramesh(2003)]%
        {ramesh2003studying}
\bibfield{author}{\bibinfo{person}{Michael Ramesh}.}
  \bibinfo{year}{2003}\natexlab{}.
\newblock \bibinfo{booktitle}{\emph{Studying public policy: Policy cycles and
  policy subsystems}}.
\newblock \bibinfo{publisher}{Don Mills, Ont.: Oxford University Press}.
\newblock


\bibitem[Riggs et~al\mbox{.}(2019)]%
        {Riggs2019-lx}
\bibfield{author}{\bibinfo{person}{William Riggs}, \bibinfo{person}{Melissa
  Ruhl}, \bibinfo{person}{Caroline Rodier}, {and} \bibinfo{person}{Will
  Baumgardner}.} \bibinfo{year}{2019}\natexlab{}.
\newblock \showarticletitle{Designing Streets for Autonomous Vehicles}. In
  \bibinfo{booktitle}{\emph{Road Vehicle Automation 6}}.
  \bibinfo{publisher}{Springer International Publishing},
  \bibinfo{pages}{111--122}.
\newblock
\urldef\tempurl%
\url{https://doi.org/10.1007/978-3-030-22933-7\_12}
\showDOI{\tempurl}


\bibitem[Rittel and Webber(1973)]%
        {wicked_problems_rittel1973}
\bibfield{author}{\bibinfo{person}{Horst W.~J. Rittel} {and}
  \bibinfo{person}{Melvin~M. Webber}.} \bibinfo{year}{1973}\natexlab{}.
\newblock \showarticletitle{Dilemmas in a General theory of planning}.
\newblock \bibinfo{journal}{\emph{Policy Sciences}} \bibinfo{volume}{4},
  \bibinfo{number}{2} (\bibinfo{date}{June} \bibinfo{year}{1973}),
  \bibinfo{pages}{155--169}.
\newblock
\showISSN{1573-0891}
\urldef\tempurl%
\url{https://doi.org/10.1007/BF01405730}
\showDOI{\tempurl}


\bibitem[Roto et~al\mbox{.}(2021)]%
        {roto2021overlaps}
\bibfield{author}{\bibinfo{person}{Virpi Roto}, \bibinfo{person}{Jung-Joo Lee},
  \bibinfo{person}{Effie Lai-Chong~Law}, {and} \bibinfo{person}{John
  Zimmerman}.} \bibinfo{year}{2021}\natexlab{}.
\newblock \showarticletitle{The overlaps and boundaries between service design
  and user experience design}. In \bibinfo{booktitle}{\emph{Designing
  Interactive Systems Conference 2021}}. \bibinfo{pages}{1915--1926}.
\newblock


\bibitem[Sandhaus et~al\mbox{.}(2023)]%
        {CHI23WS:Sandhaus_AVCityPolicy}
\bibfield{author}{\bibinfo{person}{Hauke Sandhaus}, \bibinfo{person}{Wendy Ju},
  {and} \bibinfo{person}{Qian Yang}.} \bibinfo{year}{2023}\natexlab{}.
\newblock \showarticletitle{Towards Prototyping Driverless Vehicle Behaviors,
  City Design, and Policies Simultaneously}. In
  \bibinfo{booktitle}{\emph{Workshop on Designing Technology and Policy
  Simultaneously: Towards A Research Agenda and New Practice}}.
\newblock
\showeprint[arxiv]{2304.06639}~[cs.HC]


\bibitem[Savage(2023)]%
        {Savage2023_forecasting}
\bibfield{author}{\bibinfo{person}{Tamara Savage}.}
  \bibinfo{year}{2023}\natexlab{}.
\newblock \showarticletitle{{Improving Technology Forecasting by Including
  Policy, Economic, and Social Factors}}.
\newblock  (\bibinfo{date}{9} \bibinfo{year}{2023}).
\newblock
\urldef\tempurl%
\url{https://doi.org/10.1184/R1/24103998.v1}
\showDOI{\tempurl}


\bibitem[Spaa et~al\mbox{.}(2019)]%
        {HCI-policy-boundary-CHI19}
\bibfield{author}{\bibinfo{person}{Anne Spaa}, \bibinfo{person}{Abigail
  Durrant}, \bibinfo{person}{Chris Elsden}, {and} \bibinfo{person}{John
  Vines}.} \bibinfo{year}{2019}\natexlab{}.
\newblock \bibinfo{booktitle}{\emph{Understanding the Boundaries between
  Policymaking and HCI}}.
\newblock \bibinfo{publisher}{Association for Computing Machinery},
  \bibinfo{address}{New York, NY, USA}, \bibinfo{pages}{1–15}.
\newblock
\showISBNx{9781450359702}
\urldef\tempurl%
\url{https://doi.org/10.1145/3290605.3300314}
\showURL{%
\tempurl}


\bibitem[Spaa et~al\mbox{.}(2022)]%
        {Spaa2022-evidence-and-policy}
\bibfield{author}{\bibinfo{person}{Anne Spaa}, \bibinfo{person}{Nick Spencer},
  \bibinfo{person}{Abigail Durrant}, {and} \bibinfo{person}{John Vines}.}
  \bibinfo{year}{2022}\natexlab{}.
\newblock \showarticletitle{Creative and collaborative reflective thinking to
  support policy deliberation and decision making}.
\newblock \bibinfo{journal}{\emph{Evidence \& Policy: A Journal of Research,
  Debate and Practice}}  \bibinfo{volume}{18} (\bibinfo{date}{04}
  \bibinfo{year}{2022}).
\newblock
\urldef\tempurl%
\url{https://doi.org/10.1332/174426421X16474564583952}
\showDOI{\tempurl}


\bibitem[Star and Ruhleder(1994)]%
        {star1994steps}
\bibfield{author}{\bibinfo{person}{Susan~Leigh Star} {and}
  \bibinfo{person}{Karen Ruhleder}.} \bibinfo{year}{1994}\natexlab{}.
\newblock \showarticletitle{Steps towards an ecology of infrastructure: complex
  problems in design and access for large-scale collaborative systems}. In
  \bibinfo{booktitle}{\emph{Proceedings of the 1994 ACM conference on Computer
  supported cooperative work}}. \bibinfo{pages}{253--264}.
\newblock


\bibitem[Steinhardt and Jackson(2014)]%
        {steinhardt2014reconciling}
\bibfield{author}{\bibinfo{person}{Stephanie~B Steinhardt} {and}
  \bibinfo{person}{Steven~J Jackson}.} \bibinfo{year}{2014}\natexlab{}.
\newblock \showarticletitle{Reconciling rhythms: plans and temporal alignment
  in collaborative scientific work}. In \bibinfo{booktitle}{\emph{Proceedings
  of the 17th ACM conference on Computer supported cooperative work \& social
  computing}}. \bibinfo{pages}{134--145}.
\newblock


\bibitem[Steinhardt and Jackson(2015)]%
        {steinhardt2015anticipation}
\bibfield{author}{\bibinfo{person}{Stephanie~B Steinhardt} {and}
  \bibinfo{person}{Steven~J Jackson}.} \bibinfo{year}{2015}\natexlab{}.
\newblock \showarticletitle{Anticipation work: Cultivating vision in collective
  practice}. In \bibinfo{booktitle}{\emph{Proceedings of the 18th ACM
  conference on computer supported cooperative work \& social computing}}.
  \bibinfo{pages}{443--453}.
\newblock


\bibitem[Stickdorn and Schneider(2012)]%
        {stickdorn2012service}
\bibfield{author}{\bibinfo{person}{Marc Stickdorn} {and} \bibinfo{person}{Jakob
  Schneider}.} \bibinfo{year}{2012}\natexlab{}.
\newblock \bibinfo{booktitle}{\emph{This is service design thinking: Basics,
  tools, cases}}.
\newblock \bibinfo{publisher}{John Wiley \& Sons}.
\newblock


\bibitem[The Global Partnership~of Artificial Intelligence
  Innovation~Workshop(2023)]%
        {Policybrief2023_GenAI_Jobs}
\bibfield{author}{\bibinfo{person}{Montreal The Global Partnership~of
  Artificial Intelligence Innovation~Workshop}.}
  \bibinfo{year}{2023}\natexlab{}.
\newblock \bibinfo{booktitle}{\emph{{Policy Brief: Generative AI, Jobs, and
  Policy Response}}}.
\newblock \bibinfo{type}{{T}echnical {R}eport}. \bibinfo{institution}{The
  Global Partnership of Artificial Intelligence}.
\newblock
\urldef\tempurl%
\url{https://gpai.ai/projects/future-of-work/policy-brief-generative-ai-jobs-and-policy-response-innovation-workshop-montreal-2023.pdf}
\showURL{%
\tempurl}


\bibitem[Tkacik(2020)]%
        {Tkacik_2020}
\bibfield{author}{\bibinfo{person}{Daniel Tkacik}.}
  \bibinfo{year}{2020}\natexlab{}.
\newblock \bibinfo{title}{CyLab researchers design privacy icon to be used by
  California law}.
\newblock
\newblock
\urldef\tempurl%
\url{https://www.cylab.cmu.edu/news/2020/12/11-donotsell.html}
\showURL{%
\tempurl}


\bibitem[Toxtli et~al\mbox{.}(2021)]%
        {toxtli2021quantifying}
\bibfield{author}{\bibinfo{person}{Carlos Toxtli}, \bibinfo{person}{Siddharth
  Suri}, {and} \bibinfo{person}{Saiph Savage}.}
  \bibinfo{year}{2021}\natexlab{}.
\newblock \showarticletitle{Quantifying the invisible labor in crowd work}.
\newblock \bibinfo{journal}{\emph{Proceedings of the ACM on Human-Computer
  Interaction}} \bibinfo{volume}{5}, \bibinfo{number}{CSCW2}
  (\bibinfo{year}{2021}), \bibinfo{pages}{1--26}.
\newblock


\bibitem[Urquhart and Rodden(2017)]%
        {urquhart2017new}
\bibfield{author}{\bibinfo{person}{Lachlan Urquhart} {and} \bibinfo{person}{Tom
  Rodden}.} \bibinfo{year}{2017}\natexlab{}.
\newblock \showarticletitle{New directions in information technology law:
  learning from human-computer interaction}.
\newblock \bibinfo{journal}{\emph{International Review of Law, Computers \&
  Technology}} \bibinfo{volume}{31}, \bibinfo{number}{2}
  (\bibinfo{year}{2017}), \bibinfo{pages}{150--169}.
\newblock


\bibitem[Urquhart et~al\mbox{.}(2022)]%
        {LegalProvocationAutomation_NordiCHI22}
\bibfield{author}{\bibinfo{person}{Lachlan~D. Urquhart}, \bibinfo{person}{Glenn
  McGarry}, {and} \bibinfo{person}{Andy Crabtree}.}
  \bibinfo{year}{2022}\natexlab{}.
\newblock \showarticletitle{Legal Provocations for HCI in the Design and
  Development of Trustworthy Autonomous Systems}. In
  \bibinfo{booktitle}{\emph{Nordic Human-Computer Interaction Conference}}
  (Aarhus, Denmark) \emph{(\bibinfo{series}{NordiCHI '22})}.
  \bibinfo{publisher}{Association for Computing Machinery},
  \bibinfo{address}{New York, NY, USA}, Article \bibinfo{articleno}{75},
  \bibinfo{numpages}{12}~pages.
\newblock
\showISBNx{9781450396998}
\urldef\tempurl%
\url{https://doi.org/10.1145/3546155.3546690}
\showDOI{\tempurl}


\bibitem[Users(2023)]%
        {UXRwoUsers}
\bibfield{author}{\bibinfo{person}{Synthetic Users}.}
  \bibinfo{year}{2023}\natexlab{}.
\newblock \showarticletitle{Synthetic Users: User Research without the
  Headaches}.
\newblock
\urldef\tempurl%
\url{https://www.syntheticusers.com/}
\showURL{%
\tempurl}


\bibitem[Van~Hulst and Yanow(2016)]%
        {van2016policy}
\bibfield{author}{\bibinfo{person}{Merlijn Van~Hulst} {and}
  \bibinfo{person}{Dvora Yanow}.} \bibinfo{year}{2016}\natexlab{}.
\newblock \showarticletitle{From policy “frames” to “framing”
  theorizing a more dynamic, political approach}.
\newblock \bibinfo{journal}{\emph{The American review of public
  administration}} \bibinfo{volume}{46}, \bibinfo{number}{1}
  (\bibinfo{year}{2016}), \bibinfo{pages}{92--112}.
\newblock


\bibitem[Wheeler et~al\mbox{.}(2023)]%
        {Brookings_EUAIAct_Enforcement2023}
\bibfield{author}{\bibinfo{person}{Tom Wheeler}, \bibinfo{person}{Niam~Yaraghi
  Nicol Turner~Lee}, \bibinfo{person}{John Villasenor}, {and}
  \bibinfo{person}{Nicol Turner~Lee Norman~Eisen}.}
  \bibinfo{year}{2023}\natexlab{}.
\newblock \bibinfo{title}{Key enforcement issues of the AI act should lead EU
  trilogue debate}.
\newblock
\newblock
\urldef\tempurl%
\url{https://www.brookings.edu/articles/key-enforcement-issues-of-the-ai-act-should-lead-eu-trilogue-debate/}
\showURL{%
\tempurl}


\bibitem[Whitney et~al\mbox{.}(2021)]%
        {HCITacticsForPolicyCHI21Whitney}
\bibfield{author}{\bibinfo{person}{Cedric~Deslandes Whitney},
  \bibinfo{person}{Teresa Naval}, \bibinfo{person}{Elizabeth Quepons},
  \bibinfo{person}{Simrandeep Singh}, \bibinfo{person}{Steven~R Rick}, {and}
  \bibinfo{person}{Lilly Irani}.} \bibinfo{year}{2021}\natexlab{}.
\newblock \showarticletitle{HCI Tactics for Politics from Below: Meeting the
  Challenges of Smart Cities}. In \bibinfo{booktitle}{\emph{Proceedings of the
  2021 CHI Conference on Human Factors in Computing Systems}} (Yokohama, Japan)
  \emph{(\bibinfo{series}{CHI '21})}. \bibinfo{publisher}{Association for
  Computing Machinery}, \bibinfo{address}{New York, NY, USA}, Article
  \bibinfo{articleno}{297}, \bibinfo{numpages}{15}~pages.
\newblock
\showISBNx{9781450380966}
\urldef\tempurl%
\url{https://doi.org/10.1145/3411764.3445314}
\showDOI{\tempurl}


\bibitem[Wiggers(2023)]%
        {TechCrunch2023_genAI_copoyright_lawsuits}
\bibfield{author}{\bibinfo{person}{Kyle Wiggers}.}
  \bibinfo{year}{2023}\natexlab{}.
\newblock \bibinfo{title}{The current legal cases against Generative AI are
  just the beginning}.
\newblock
\newblock
\urldef\tempurl%
\url{https://techcrunch.com/2023/01/27/the-current-legal-cases-against-generative-ai-are-just-the-beginning/}
\showURL{%
\tempurl}


\bibitem[Wong and Jackson(2015)]%
        {wong2015wireless}
\bibfield{author}{\bibinfo{person}{Richmond~Y Wong} {and}
  \bibinfo{person}{Steven~J Jackson}.} \bibinfo{year}{2015}\natexlab{}.
\newblock \showarticletitle{Wireless visions: Infrastructure, imagination, and
  US spectrum policy}. In \bibinfo{booktitle}{\emph{Proceedings of the 18th ACM
  Conference on Computer Supported Cooperative Work \& Social Computing}}.
  \bibinfo{pages}{105--115}.
\newblock


\bibitem[Wong and Mulligan(2019)]%
        {wong2019bringing}
\bibfield{author}{\bibinfo{person}{Richmond~Y Wong} {and}
  \bibinfo{person}{Deirdre~K Mulligan}.} \bibinfo{year}{2019}\natexlab{}.
\newblock \showarticletitle{Bringing design to the privacy table: Broadening
  “design” in “privacy by design” through the lens of HCI}. In
  \bibinfo{booktitle}{\emph{Proceedings of the 2019 CHI conference on human
  factors in computing systems}}. \bibinfo{pages}{1--17}.
\newblock


\bibitem[Yang et~al\mbox{.}(2020)]%
        {yangUXMLframework}
\bibfield{author}{\bibinfo{person}{Qian Yang}, \bibinfo{person}{Aaron
  Steinfeld}, \bibinfo{person}{Carolyn Ros\'{e}}, {and} \bibinfo{person}{John
  Zimmerman}.} \bibinfo{year}{2020}\natexlab{}.
\newblock \showarticletitle{Re-Examining Whether, Why, and How Human-AI
  Interaction Is Uniquely Difficult to Design}. In
  \bibinfo{booktitle}{\emph{Proceedings of the 2020 CHI Conference on Human
  Factors in Computing Systems}} (Honolulu, HI, USA)
  \emph{(\bibinfo{series}{CHI '20})}. \bibinfo{publisher}{Association for
  Computing Machinery}, \bibinfo{address}{New York, NY, USA},
  \bibinfo{pages}{1–13}.
\newblock
\showISBNx{9781450367080}
\urldef\tempurl%
\url{https://doi.org/10.1145/3313831.3376301}
\showDOI{\tempurl}


\bibitem[Yang et~al\mbox{.}(2023)]%
        {design-policy-workshop}
\bibfield{author}{\bibinfo{person}{Qian Yang}, \bibinfo{person}{Richmond~Y.
  Wong}, \bibinfo{person}{Thomas Gilbert}, \bibinfo{person}{Margaret~D. Hagan},
  \bibinfo{person}{Steven Jackson}, \bibinfo{person}{Sabine Junginger}, {and}
  \bibinfo{person}{John Zimmerman}.} \bibinfo{year}{2023}\natexlab{}.
\newblock \bibinfo{booktitle}{\emph{Designing Technology and Policy
  Simultaneously: Towards A Research Agenda and New Practice}}.
\newblock \bibinfo{publisher}{Association for Computing Machinery},
  \bibinfo{address}{New York, NY, USA}.
\newblock
\showISBNx{9781450394222}
\urldef\tempurl%
\url{https://doi.org/10.1145/3544549.3573827}
\showDOI{\tempurl}


\bibitem[Young et~al\mbox{.}(2019)]%
        {young2019toward}
\bibfield{author}{\bibinfo{person}{Meg Young}, \bibinfo{person}{Lassana
  Magassa}, {and} \bibinfo{person}{Batya Friedman}.}
  \bibinfo{year}{2019}\natexlab{}.
\newblock \showarticletitle{Toward inclusive tech policy design: a method for
  underrepresented voices to strengthen tech policy documents}.
\newblock \bibinfo{journal}{\emph{Ethics and Information Technology}}
  \bibinfo{volume}{21} (\bibinfo{year}{2019}), \bibinfo{pages}{89--103}.
\newblock


\bibitem[Zamfirescu-Pereira et~al\mbox{.}(2023)]%
        {zamfirescu2023johnny}
\bibfield{author}{\bibinfo{person}{JD Zamfirescu-Pereira},
  \bibinfo{person}{Richmond~Y Wong}, \bibinfo{person}{Bjoern Hartmann}, {and}
  \bibinfo{person}{Qian Yang}.} \bibinfo{year}{2023}\natexlab{}.
\newblock \showarticletitle{Why Johnny can’t prompt: how non-AI experts try
  (and fail) to design LLM prompts}. In \bibinfo{booktitle}{\emph{Proceedings
  of the 2023 CHI Conference on Human Factors in Computing Systems}}.
  \bibinfo{pages}{1--21}.
\newblock


\end{thebibliography}
\balance
\appendix
\section*{Appendix}

\section{Data Analysis Process}\label{appendix:analysisMethod}
We started by proposing many possible problem-solution frames that meet the following three criteria. Next, we iteratively critiqued and improved these emergent frames, also based on these criteria.

\begin{itemize}[leftmargin=*]
    \item \textit{Effective problem framing:} It can help explain the disparate success levels observed in HCI-policy collaborations so far;
    \item \textit{Useful solution framing:} It can reveal novel strategies for addressing the disparity;
    \item \textit{Adaptive solution framing:} The strategies it reveals are flexible, allowing HCI researchers and practitioners to derive specific actions in diverse situations. This flexibility is crucial, because effective approaches to HCI-policy collaboration vary based on contexts~\cite{haase2019meaning}. Our goal is not to prescribe fixed methods, but to identify new \textit{avenues} of improvements. 
\end{itemize}

Besides meeting the above criteria, the final frame is also a useful seed for HCI-community-wide discussions. 
Even the authors lack consensus on whether this problem-solution frame might be too progressive and controversial, or too obvious that it represents what HCI needs to do. In this sense, we see this frame as an invitation (and potentially provocation) for fellow HCI researchers, practitioners, and educators to debate the nature of HCI-policy collaboration challenges. We encourage them to share their frames and collectively deliberate how to best move forward.

\section{Why Interactive, Infrastructure Technologies Necessitate Deep, Continuous HCI-Policy Partnership}
\label{appendix:infraTech}

AI Foundation Models (FMs) such as chatGPT are telling examples. FMs are machine learning models that AI engineers and end users alike can easily adapt and modify to create bespoke text-, image-, or video-generation applications~\cite{zamfirescu2023johnny,bommasani2021opportunities}. 
Without policy expertise, HCI's futuring and empirical methods cannot effectively understand FMs' societal impact or collect empirical evidence proving this impact.

\begin{itemize}[leftmargin=*]
    \item \textbf{Each FM powers an AI-model-and-app ecosystem. HCI cannot understand the FM's full impact without considering the policies that govern it.} For example, chatGPT. Its societal impact depends on the various fine-tuned models and downstream apps people build with it~\cite{CriminalGPT_Wired2023}. Merely empirically studying how one model (e.g., chatGPT, August 8th, 2023 version) impacts one user here and now misses the forest for the trees. Instead, HCI empirical work must also understand how policies incentivize and regulate the chatGPT-derived-model-and-app ecosystem to understand its societal impact (the “forest.”)
    
    This need for policy considerations is not a coincidence. From sensor networks covering a smart city to algorithm-mediated social media platforms, infrastructure technologies operate at societal-level scales and power an ecosystem of various downstream applications. To understand their societal impact, one must first understand the policies that govern the ecosystem.

    \item \textbf{HCI cannot prove the relationship between FM design choices and  societal impact, without HCI and policy join forces}. 
    Companies like OpenAI, which own FMs and their ecosystems, hold enormous power over who can know which aspects of FMs' inner workings and data practices. However, this information is critical for studying FM's legality and social impact. Without the joint forces of policy mandates and HCI knowledge (e.g., on what information is critical for understanding FM's fairness), these companies may never provide the public \textit{meaningful} access to FMs' inner workings.

    This need for policy mandates on technology transparency is not coincidental either. The effect of cyberinfrastructure on people is often indirect and invisible. Understanding this effect requires information about its inner workings. However, organizations that own infrastructure technologies (e.g., social media giants, governments) are few and powerful, holding enormous power in guarding this information. In this context, HCI needs policy's power and policy expertise, which can enforce information disclosure and know how to do so without harming tech industry competition.

    \item \textbf{Given that FMs' societal impact is far-reaching and demands quick regulation, empirical researchers need to analyze their societal impact with policymakers side-by-side.} Unlike in some previous cases, HCI does not seem to have the luxury of time to amass empirical evidence of FMs' societal pros and cons before policymakers demand them~\cite{chatGPT_quickRegulation}.
    
    The fact that FMs seem to have impacted “everything everywhere all at once” is not coincidental. From Uber (mobile computing combined with gig work) to social media feed-ranking algorithms, infrastructure technologies, once established and adopted at scale, can directly impact human and human-computer interactions of all sorts. Understanding their societal impact requires HCI and policy to collaborate synchronously. Moreover, they need to develop such understanding within specific windows of opportunity, for example, before the technology is widely adopted and its associated human practices settle in.
    
\end{itemize}

For HCI designers who innovate human-centered systems, policy considerations are also crucial from the outset. 
\begin{itemize}[leftmargin=*]
    \item \textbf{If the FM violates the law, HCI’s novel FM apps may also violate the law. }By creating one accessible system design, HCI incrementally improves web accessibility. Creating one new GPT app without policy considerations, however, HCI risks violating copyright laws. 

    \item \textbf{FMs' unprecedented malleability means that HCI's well-intentioned app designs can easily be misused.~}It seems naive to think an FM tool that predicts the likely outcomes of a user study protocol would not have been used to power ``\textit{user studies without users}'' \cite{UXRwoUsers}, or an FM tool that helps Reddit moderators to predict upcoming fake news would not have been used to generate fake news instead. Without policy guardrails to regulate bad actors, HCI's well-intentioned app designs can play an unintended role in FMs' societal harms.

These seemingly unusual risks of HCI designing illegal or unethical things are not coincidental. Just like VR platforms transformed how governments offer public services \cite{AccentureVRgov_2023} and GPT changed how HCI software designers design apps, infrastructure technologies change how almost everyone---including HCI researchers, practitioners, policymakers, and policy enforcement agencies---works. These changes complicate HCI's efforts to improve or regulate infrastructure technologies' societal impact, offering yet another reason for HCI to collaborate with policy experts.

\end{itemize} 

\section{Responding to Regulatory Agencies' Requests for Comment}
In the U.S., regulatory agencies (they are responsible for creating rules that enact laws into practice, such as the FTC or National Institute of Standards and Technology) continually seek input and feedback from both experts and the general public, often through requests for public comments for 30-60 days, allowing the public to comment or submit data related to proposed rules or decision-making \cite{FederalRegister_rulemaking}. 
HCI professionals can submit empirical evidence of tech's human impact as expert evidence. 
This evidence could inform policies that both regulate (or promote) certain forms of technology development and use, or that regulate (or promote certain human behaviors related to technology use. 

\balance

\end{document}